\documentclass[prl,aps,twocolumn,showpacs,10pt]{revtex4-1}

\usepackage{graphicx}
\usepackage{dcolumn}
\usepackage{bm}
\usepackage{amssymb,amsmath}
\usepackage{hyperref}
\usepackage{txfonts}

\usepackage[usenames,dvipsnames]{color}

\newcommand{\sub}[1]{_{\text{#1}}}

\newcommand{\Er}[1]{Equation~\eqref{#1}}
\newcommand{\er}[1]{Eq.~\eqref{#1}}

\newcommand{\reffig}[1]{Fig.~\ref{#1}}
\newcommand{\punc}[1]{\,\text{#1}}

\newcommand{\Phiv}{\bm{\Phi}}
\newcommand{\rv}{{\bm{r}}}

\usepackage{graphicx}
\usepackage{leftidx}

\newcommand{\ket}[1]{\left| #1 \right>} 
\newcommand{\bra}[1]{\left< #1 \right|}

\begin{document}

\title{Quantum Slow Relaxation and Metastability due to Dynamical Constraints}
\author{Zhihao Lan}
\affiliation{Centre for the Mathematics and Theoretical Physics of Quantum Non-equilibrium Systems} 
\affiliation{School of Physics and Astronomy, University of Nottingham, Nottingham NG7 2RD, United Kingdom}
\author{Merlijn van Horssen}
\affiliation{Centre for the Mathematics and Theoretical Physics of Quantum Non-equilibrium Systems} 
\affiliation{School of Physics and Astronomy, University of Nottingham, Nottingham NG7 2RD, United Kingdom}
\author{Stephen Powell}
\affiliation{Centre for the Mathematics and Theoretical Physics of Quantum Non-equilibrium Systems} 
\affiliation{School of Physics and Astronomy, University of Nottingham, Nottingham NG7 2RD, United Kingdom}
\author{Juan P. Garrahan}
\affiliation{Centre for the Mathematics and Theoretical Physics of Quantum Non-equilibrium Systems} 
\affiliation{School of Physics and Astronomy, University of Nottingham, Nottingham NG7 2RD, United Kingdom}

\begin{abstract}
One of the general mechanisms that give rise to the slow cooperative relaxation characteristic of classical glasses is the presence of kinetic constraints in the dynamics.  Here we show that dynamical constraints can similarly lead to slow thermalization and metastability in translationally invariant quantum many-body systems.  We illustrate this general idea by considering two simple models: (i) a one-dimensional quantum analogue to classical constrained lattice gases where excitation hopping is constrained by the state of neighboring sites, mimicking excluded-volume interactions of dense fluids; and (ii) fully packed quantum dimers on the square lattice.  Both models have a Rokhsar--Kivelson (RK) point at which kinetic and potential energy constants are equal.  To one side of the RK point, where kinetic energy dominates, thermalization is fast.  To the other, where potential energy dominates, thermalization is slow, memory of initial conditions persists for long times, and separation of timescales leads to pronounced metastability before eventual thermalization.  Furthermore, in analogy with what occurs in the relaxation of classical glasses, the slow-thermalization regime displays dynamical heterogeneity as manifested by spatially segregated growth of entanglement. 
\end{abstract}
\maketitle

\noindent
{\bf \em Introduction.}---Interacting quantum systems generically {\em equilibrate}: their long-time state after unitary evolution under the Hamiltonian is, loosely speaking, indistinguishable from the time-integrated state, as concerns expectation values of local observables  \cite{Reimann2008,Linden2009,Short2011,Short2012,Reimann2012}.  Equilibration requires (almost) no degeneracies in energy gaps and stationarity is due to dephasing in the energy eigenbasis  \cite{Gogolin2016,DAlessio2016,Borgonovi2016}. 
Most quantum many-body systems, furthermore, are believed to {\em thermalize} \cite{Gogolin2016,DAlessio2016,Borgonovi2016}: if $A$ and $B$ are partitions, the reduced state in $A$ at long times tends to ${\rm Tr}_B[e^{-\beta H}]$, with temperature $1/\beta$ set by the constant $\langle H\rangle$ \cite{Gogolin2016,DAlessio2016,Borgonovi2016}.  
Expectation values in $A$ hence take thermal values, and memory of initial conditions is lost except for the energy.  This is the general setup for quantum ergodicity, where the system acts as its own thermal reservoir \cite{Gogolin2016,DAlessio2016,Borgonovi2016}. 
Thermalization can be seen as a consequence of the eigenstate thermalization hypothesis (ETH) \cite{Deutsch1991,Srednicki1994,Tasaki1998,Rigol2008}.  

Exceptions to this scenario include integrable systems \cite{Essler2016} which equilibrate to a generalized  Gibbs ensemble (i.e., being ``as ergodic as possible'' given their large number of conserved quantities) \cite{Rigol2007,Vidmar2016}. Another notable exception is {\em many-body localization} (MBL)~\cite{Altshuler1997,Basko2006,Gornyi2005,Oganesyan2007,Znidaric2008,Pal2010,Bardarson2012,Serbyn2013,Huse2014,Andraschko2014,Yao2014,Serbyn2014,Laumann2014,De-Roeck2014,Scardicchio2015,Vasseur2015,Agarwal2015,Bar-Lev2015,Imbrie2016,Schreiber2015,Bordia2016,Smith2016} displayed by many-body quantum systems with quenched disorder; for reviews see \cite{Nandkishore2015,Altman2015,Abanin2017}. Under MBL conditions -- typically when the disorder exceeds some threshold -- ETH breaks down, dynamics becomes nonergodic, and the long-time state depends on initial conditions.

\begin{figure}
	\includegraphics[width=\columnwidth]{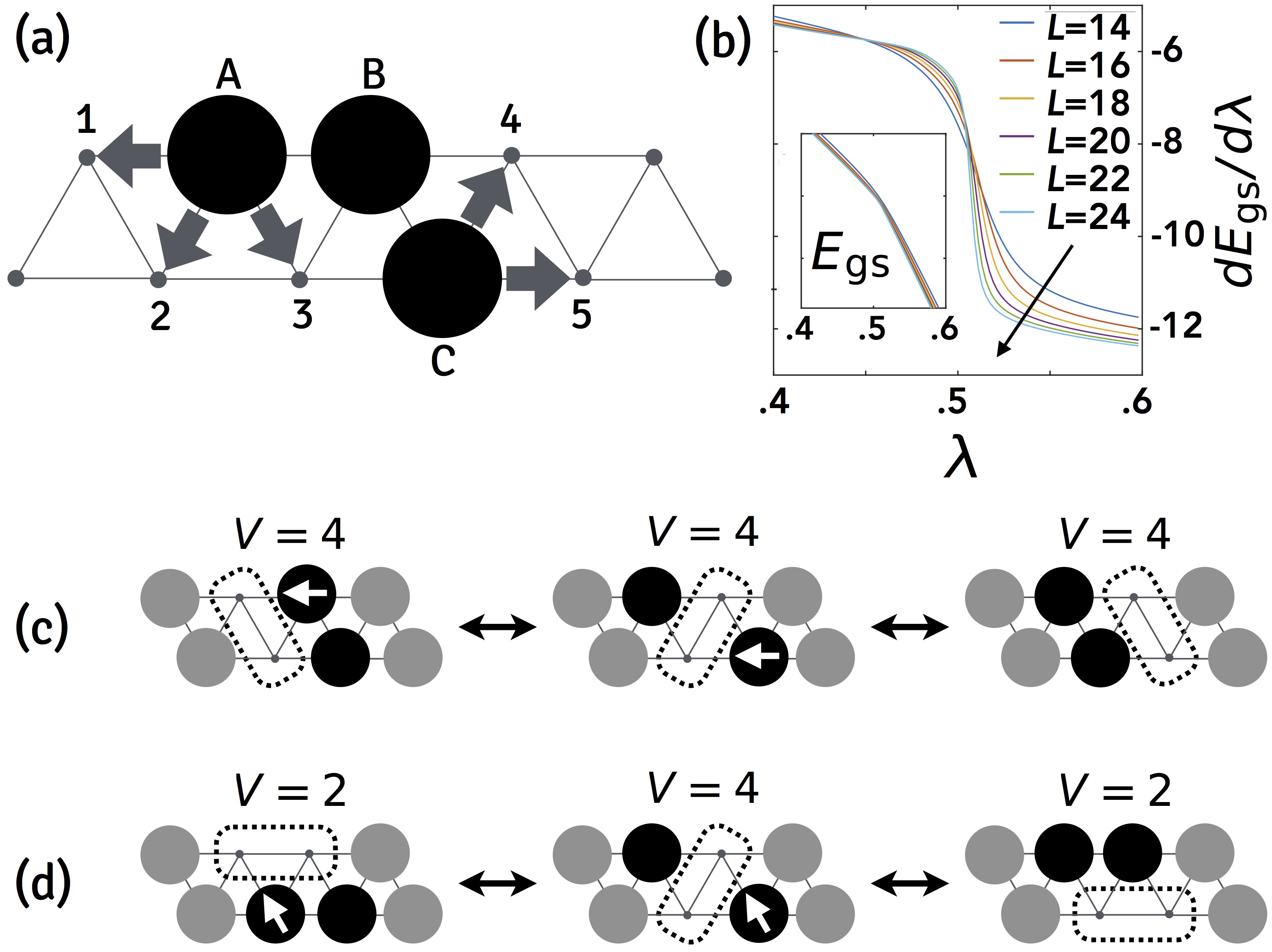}
\caption{Constrained 1D quantum lattice gas. (a) Particle hops, with amplitude \(\lambda\), are only allowed if at least one common neighbor of the initial and final sites is empty (arrows indicate allowed moves).  For example, C can hop to both 4 and 5, but not to 3 due to B. Each link for which the constraint is satisfied gives an interaction energy $1-\lambda$. (b) Quantum phase transition at the RK point: the ground state energy (Inset) has a first-order singularity at $\lambda = 1/2$ for large $L$ (filling fractions $N/L$ with $L-N = 4$). (c),(d) Effective hopping of vacancy dimers, indicating the potential energy (in units of \(1-\lambda\)) of each configuration. }
\label{fig1}
\end{figure}

One can compare the above to mechanisms for classical nonergodicity.  MBL is analogous to classical systems with random fields or interactions, such as spin glasses \cite{Binder1986}, where strong disorder leads to thermodynamic phase transitions to nonergodic states.  But classically, disorder is not the only mechanism that impedes relaxation.  Structural glasses, such as those formed from supercooled liquids or densified colloids, are nonthermalising {\em without} quenched disorder \cite{Binder2011,Berthier2011,Biroli2013}.  The central ingredients are excluded-volume (steric) interactions that lead to effective {\em kinetic constraints} in the dynamics \cite{Fredrickson1984,Palmer1984,Ritort2003}.  In contrast to spin-glasses, it is debated \cite{Lubchenko2007,Chandler2010,Binder2011,Berthier2011,Biroli2013} whether structural glasses eventually undergo a phase transition to a truly nonergodic state, or if, given enough time, they would  eventually thermalize.  If the latter, they are dynamically {\em metastable}, appearing nonergodic on experimental timescales.  Similarly, an important open question in quantum nonergodicity is whether MBL is possible in translational invariant systems 
\cite{Carleo2012,Horssen2015,Schiulaz2015,Papic2015,Barbiero2015,Yao2016,Prem2017,Smith2017,Yarloo2018,Mondaini2017}.  

Here we address the question of slow quantum relaxation in nondisordered systems due to dynamical constraints. We consider systems that obey ETH -- and thus thermalize asymptotically -- but where thermalization is slow due to a separation of timescales that leads to pronounced metastability.  We consider two prototypical models, a one-dimensional (1D) quantum analogue to classical constrained lattice gases \cite{Kob1993,Jackle1994,Ritort2003,Pan2005} and quantum dimers on the two-dimensional (2D) square lattice \cite{Rokhsar1988,Moessner2011,Chalker2017}.  In both cases, we show the existence of slow relaxing regimes when interactions dominate over kinetic energy.  As in classical glasses, we find that metastability is associated to spatially heterogeneous relaxation dynamics.

%%%%%%%%%%%%%%%%%%%%%%%%%%
\noindent
{\bf \em 1D constrained quantum lattice gas.}---Consider hard-core particles moving on a 1D strip of a triangular lattice with $L$ sites (and periodic boundary conditions along the strip) and $N$ particles; see \reffig{fig1}. The Hamiltonian is
\begin{multline} \label{H}
	H\sub{QLG} = -\tfrac{1}{2} \sum_{\langle i, j\rangle} 
	\hat{C}_{ij} \left\{
	\lambda
	\left( 
	\sigma^{+}_{i} \sigma^{-}_{j} + \sigma^{+}_{j} \sigma^{-}_{i}
	\right) 
	\right.
\\
	\left.
	- (1-\lambda) 
	\left[
	n_{i} (1 - n_{j}) + n_{j}(1 - n_{i})
	\right] 
	\right\}
	\punc.
\end{multline}
Here $\sigma^{+}_{i} = \vert 1_{i} \rangle\langle 0_{i} \vert$, $\sigma^{-}_{i} = \vert 0_{i} \rangle \langle 1_{i} \vert$, $n_{i} = \sigma^{+}_{i} \sigma^{-}_{i}$, with $\vert 0_{i} \rangle $ and $\vert 1_{i} \rangle$ the empty and occupied states on site $i$, respectively, and the sum is over nearest neighbors ${\langle i, j\rangle}$.   The operator $\hat{C}_{ij} = 1 - \prod_k n_k$ is a {\em dynamical constraint}, where the product is over all common-neighbor sites $k$ of $i$ and $j$. As for classical constrained lattice gases \cite{Kob1993,Jackle1994,Ritort2003,Pan2005}, $\hat{C}_{ij}$ mimics steric restrictions: particles occupy finite volume and impede motion of their neighbors; see \reffig{fig1}(a). The model conserves density but has no particle--hole symmetry.  The effect of the constraints is only significant for large fillings, where many moves possible in the unconstrained problem are blocked.

The first term of the summand in \er{H} describes nearest-neighbor hopping with frequency $\lambda$, while the second is an interaction energy between the same neighbors of strength $1-\lambda$.  Both terms vanish if the constraint on the bond is not satisfied, and thus, only bonds for which $\hat{C}_{ij} \neq 0$ contribute
\footnote{
The structure of $H\sub{QLG}$ is similar to those in \cite{Horssen2015} and \cite{Hickey2016}.  Constraints partition Hilbert space into disconnected components: states with only isolated vacancies cannot be dynamically connected with $H\sub{QLG}$; but most states have at least one pair of neighouring vacancies and belong to the ergodic partition (we consider the dynamics in this main subspace).  The model here and those of \cite{Horssen2015,Hickey2016} are termed ``embedded'' Hamiltonians in \cite{Shiraishi2017}.
}.
The system has a Rokhsar--Kivelson (RK) point at $\lambda = 1/2$ \cite{Rokhsar1988,Castelnovo2005}: the Hamiltonian is equivalent to (minus) the generator of classical stochastic dynamics and the ground-state wave function is given by an equal superposition of all classical states for each filling.  For $0 < \lambda \neq 1/2$, $H\sub{QLG}$ is also related to classical dynamics, being (minus) the ``tilted'' generator for ensembles of trajectories whose probability is biased by $[\lambda/(1-\lambda)]^K$ with $K$ the total number of particle hops \cite{Lecomte2007,Garrahan2009}.  The ground-state energy of $H\sub{QLG}$ then gives the large-deviation \cite{Touchette2009} cumulant-generating function of $K$.  For constrained lattice gases, it is known \cite{Garrahan2009} that this has a first-order singularity at $\lambda = 1/2$ in the large size limit, corresponding to a quantum phase transition in the quantum problem; see \reffig{fig1}(b).

\begin{figure}
	\includegraphics[width=\columnwidth]{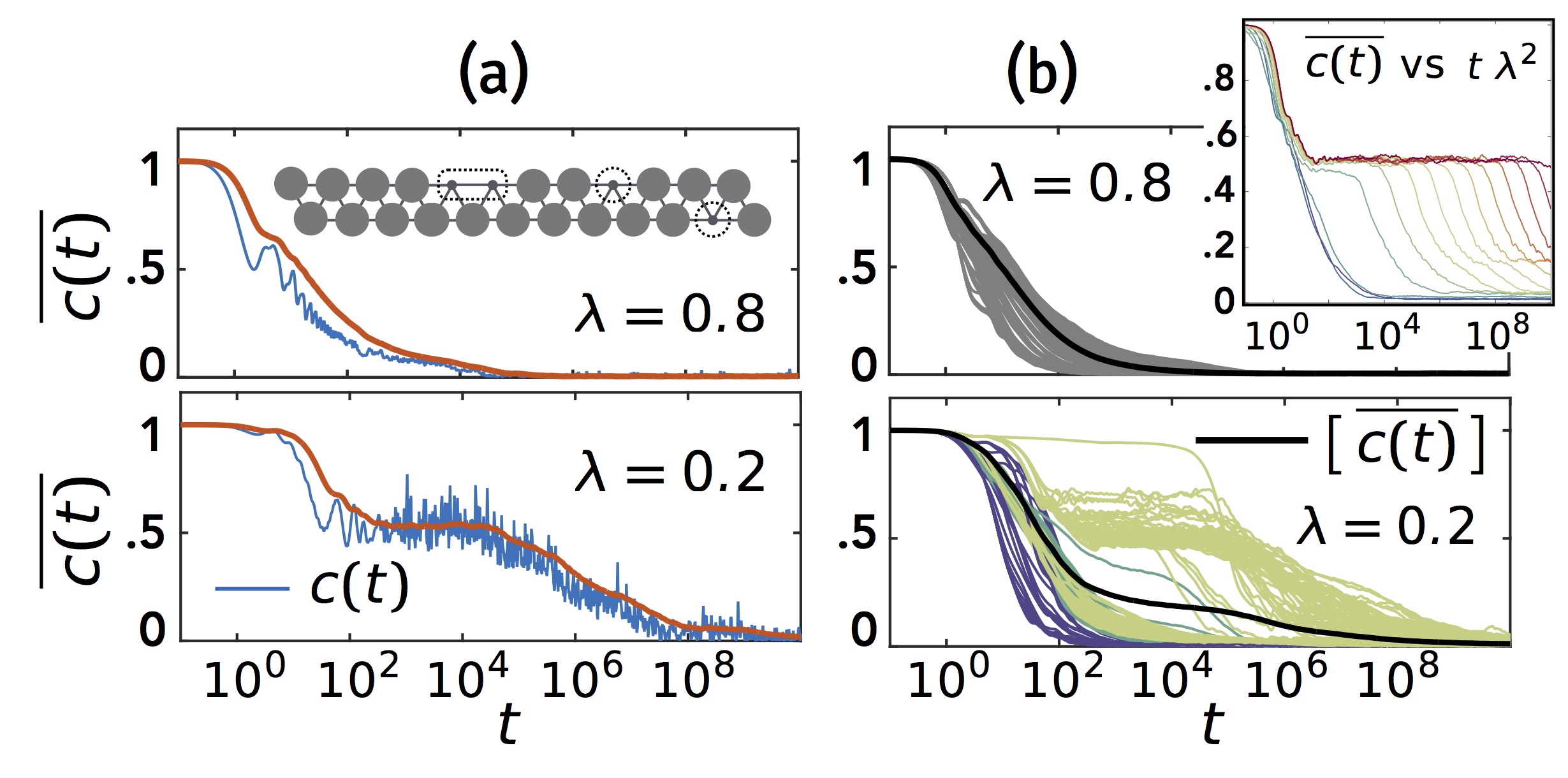}
\caption{(a) Decay of the normalized density autocorrelator with time, for $\lambda = 0.8$ (top) and $\lambda = 0.2$ (bottom).  The blue curve is $c(t)$ and the orange one $\overline{c(t)}$. The inset shows the initial configuration, with $L=24$ and $N=20$.  (b) Density correlations for all product initial states.  The thick black curve corresponds to the $T=\infty$ average, $[\overline{c(t)}]$, over initial states at this filling ($L=24$, $N=20$).
Inset: autocorrelations for the initial state of (a) for various $\lambda$ versus rescaled time $t \lambda^2$. 
}
\label{fig2}
\end{figure}

\begin{figure*}
	\includegraphics[width=\textwidth]{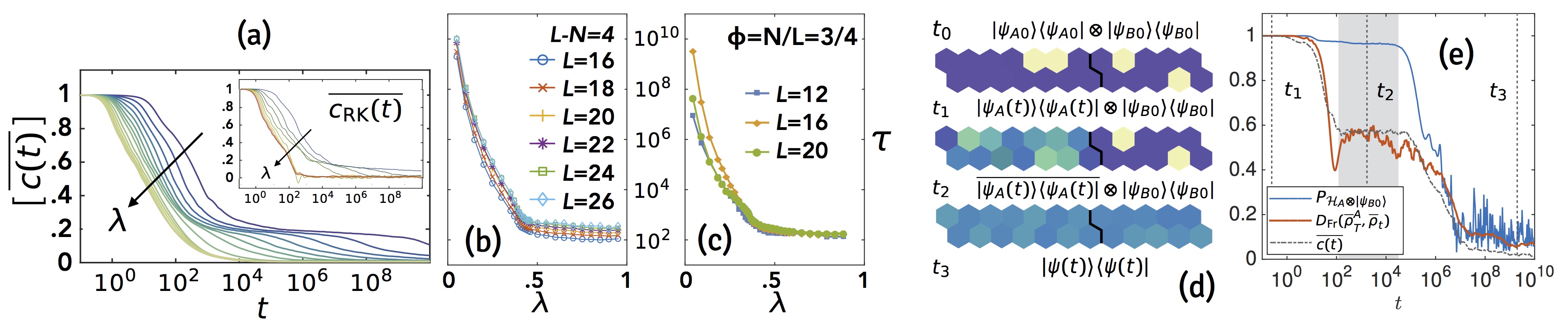}
\caption{(a) $T=\infty$ average, time-averaged density autocorrelation, $[\overline{c(t)}]$, for varying $\lambda$. Inset: same but with the RK ground state as initial condition, $\overline{c_{\rm RK}(t)}$. The relaxation behavior is similar to the $T=\infty$ average despite the fact that the RK state is entangled. (b) Relaxation time $\tau$ extracted from the average $T=\infty$ correlators, as a function of $\lambda$ for the sizes shown for {\em fixed number of vacancies} $L-N=4$. (c) Same for {\em fixed filling fraction} $\phi = N/L = 3/4$. For the small sizes accessible to numerics, there is a small dependence on the parity of $N$.
(d) Dynamically heterogeneous relaxation: average local occupation in the three time regimes of $c(t)$ starting from the initial configuration of \reffig{fig2}(a). (For movies see \cite{SM}.)
(e) Weight of the projection of $\rho(t)$ onto the subspace $\mathcal{H}_{A} \otimes \vert \psi_{B0} \rangle$ (blue) showing that regimes $t_1$ and $t_2$ correspond to growth of entanglement in $A$ only.  The distance between the full time-integrated states $\overline{\rho(t)}$ and $\overline{\rho(t)}^A = \overline{| \psi_{A}(t) \rangle \langle \psi_{A}(t) |}
 \otimes 
| \psi_{B0} \rangle \langle \psi_{B0} |$ tracks closely the evolution of $c(t)$, as seen from 
the (normalized) Frobenius norm, $D_{\mathrm{Fr}}(\rho,\sigma) = \sqrt{ \mathrm{Tr} \left[ (\rho - \sigma)^{2} \right]} / \sqrt{ \mathrm{Tr} \left[\rho^{2}\right] + \mathrm{Tr} \left[\sigma^{2}\right] }$ (orange).
}
\label{fig3}
\end{figure*}

We consider evolution under the dynamics generated by \er{H}, $\ket{\psi(t)} =  e^{-i H\sub{QLG} t} \ket{\psi_0}$, taking as initial states $\ket{\psi_0}$ product states corresponding to classical configurations, (discarding those with only isolated vacancies, which are disconnected under $H\sub{QLG}$). To quantify relaxation, we study two-time correlation functions, in particular the autocorrelator,
\begin{equation}
c(t) = \frac{1}{L} \sum_i 
\frac{\langle \psi_0 | n_i(t) n_i(0) |\psi_0\rangle}
{\phi(1-\phi)}
-
\frac{\phi}{(1-\phi)} 
\label{c}
\punc,
\end{equation}
where \(n_i(t)\) is the Heisenberg-picture number operator and $\phi=N/L$ is the filling fraction.  \Er{c} defines the connected correlator, scaled to go from $c(0)=1$ to $c(\infty)=0$. Since $|\psi_0 \rangle$ is a product state, $\langle \psi_0 | n_i(t) n_i(0) |\psi_0\rangle$ reduces to the expectation value $\langle n_i(t) \rangle$ for initially occupied sites $i$.  

Figure 2(a) shows $c(t)$ and the time-averaged $\overline{c(t)} = t^{-1} \int_0^t dt' c(t')$ (to smooth out short-scale fluctuations) for one particular initial condition.  For $\lambda = 0.8$, the kinetic term in $H\sub{QLG}$ dominates over the potential and thermalization is fast.  In sharp contrast, for $\lambda = 0.2$, where potential energy dominates over kinetic, $\overline{c(t)}$ displays a pronounced separation of timescales, decaying fast to a nonzero plateau, and thermalizing only at much longer times. Such two-step correlators are typical of classical glassy systems \cite{Binder2011,Berthier2011,Biroli2013}.  Figure 2(b) shows $\overline{c(t)}$ for all product-state initial conditions.  For $\lambda > 1/2$, there is little variation between initial conditions, and all correlators decay rapidly.  In turn, for $\lambda < 1/2$, there is a strong dependence on initial conditions, some thermalizing fast, while others thermalize much more slowly.

This can be understood as follows.  For small $\lambda$, we can consider the hopping term in $H\sub{QLG}$ perturbatively.  The simplest mechanism for relaxation is effective hopping of dimers of vacancies, cf.\ \reffig{fig1}(c,d), which requires the hybridization of unperturbed states with energy $V$.  Dimers therefore diffuse with an effective rate scaling as $\lambda^2$.  However, when a dimer encounters an isolated vacancy, this mechanism breaks down as the corresponding states become off-resonant; isolated vacancies therefore act as barriers to dimer propagation.  The separation of timescales can be seen in the inset of \reffig{fig2}(b), which shows $\overline{c(t)}$ for the initial state of \reffig{fig2}(a) for varying $\lambda$: the rate $\lambda^2$ accounts for the whole correlators in the fast regime ($\lambda > 1/2$) but only up to the plateau in the slow regime ($\lambda < 1/2$) where subsequent relaxation requires more complex collective processes. 

Figure \ref{fig3}(a) shows the autocorrelator for an equal mixture of all initial conditions (infinite-temperature average), $[\overline{c(t)}]$. It is dominated by slow-relaxing initial states [i.e., those with isolated vacancies, cf.,\ inset of \reffig{fig2}(a)] and displays two-step behavior for $\lambda < 1/2$. The inset to   
Fig.~\ref{fig3}(a) shows the (time-averaged) autocorrelator $\overline{c_{\rm RK}(t)}$ for an initial state that is the ground state at the RK point $\lambda = 1/2$ (an equal superposition of all basis states), amounting to a quench from the RK point. In contrast to the product states of the $T=\infty$ mixture, this initial state is entangled. Nonetheless, slow relaxation for $\lambda < 1/2$ is still apparent.

An overall relaxation time $\tau$ can be defined from $[\overline{c(\tau)}] = \epsilon$.  The values of $\tau$ for a threshold $\epsilon = 10^{-1}$ are shown in \reffig{fig3}(b,c) as a function of $\lambda$: in (b) we fix the number of vacancies $L-N$ and change system size $L$, while in (c) we fix the filling $\phi = N/L$. 
In both cases, there is a clear change around the RK point, $\lambda = 1/2$, from a regime where the timescale grows modestly, to one where $\tau$ increases substantially with decreasing $\lambda$. 
In particular from \reffig{fig3}(c), we expect that this behavior will persist in the limit $L,N \to \infty$ with $\phi$ fixed.

\begin{figure}
	\includegraphics[width=\columnwidth]{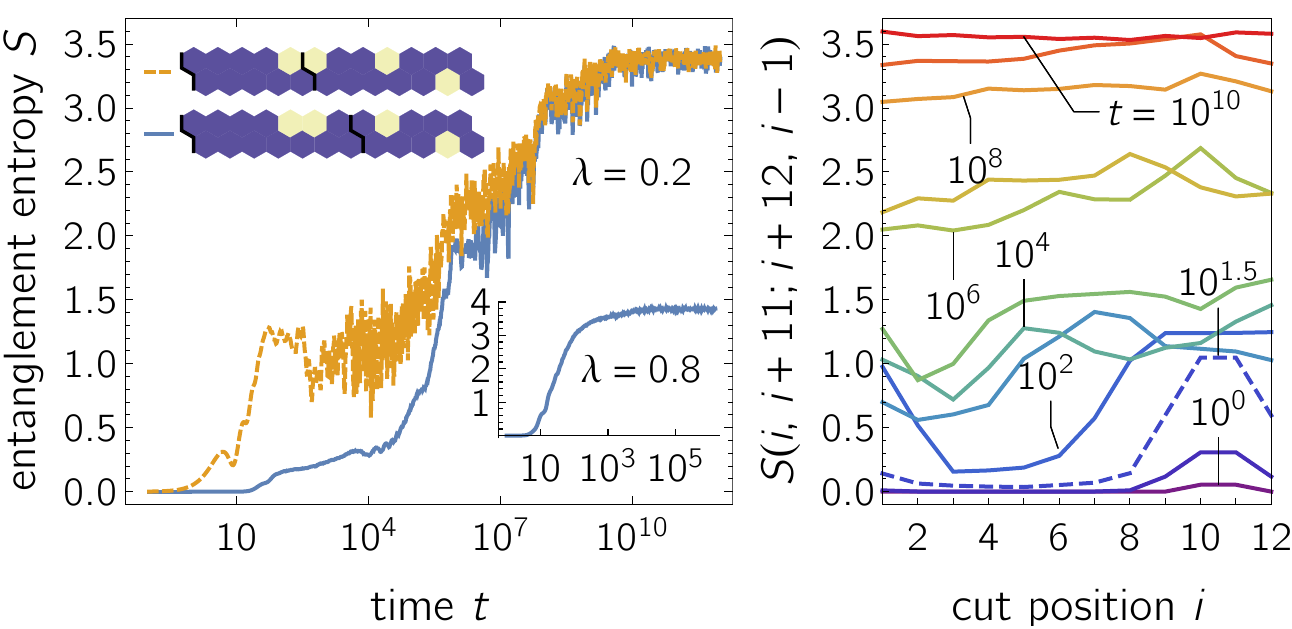}
\caption{Spatial heterogeneity of entanglement. Left: Entanglement entropy for two choices of partition, shown in upper inset, for dynamics starting from configuration shown, for \(\lambda=0.2\) and (inset) \(\lambda=0.8\). Right: Entanglement entropy as a function of the location of the partition at various times \(t\). The two partitions consist of, respectively, sites \(i\) to \(i + 11\) and the complement, \(i+12\) to \(i-1\) (with periodic boundaries). Labels show \(t\); going upwards, each successive line has \(t\) increased by a factor of \(10\), except the dashed line, which has \(t = 10^{1.5}\).}
\label{EntanglementEntropyPlot}
\end{figure}

\begin{figure*}
\centering
\includegraphics[width=\textwidth]{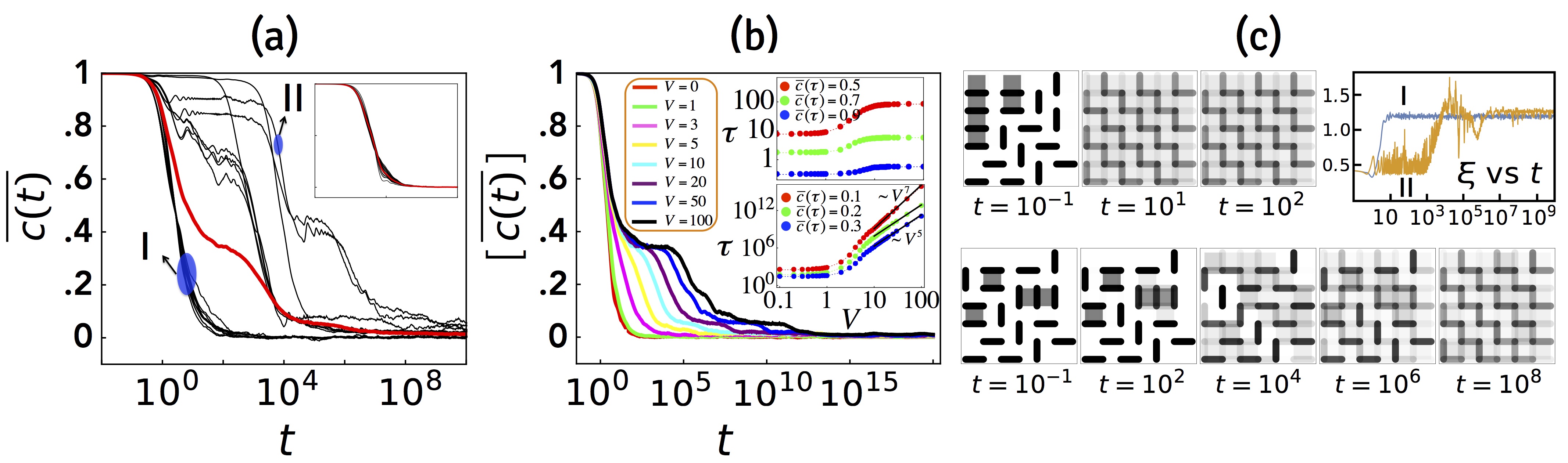}
\caption{Relaxation dynamics of the QDM on a $6\times6$ square lattice. (a) Normalized two-time dimer correlation $\overline{c(t)}$ for different initial configurations in the \((1,1)\) flux sector. 
Relaxation is slow for \(V=10\) but fast for \(V =0.5\) (inset). The red curves show the $T=\infty$ average, $[\overline{c(t)}]$.
(b) $[\overline{c(t)}]$ for various \(V\). Insets: times \(\tau\) at which certain thresholds of $\overline{c(t)}$  are reached, versus \(V\). (c) Spatial distribution of potential energy (plaquette flippability) as a function of time at $V=10$, starting from two different dimer configurations, labeled I and II in (a), with fast and slow relaxation, respectively (for movies see \cite{SM}). For II, remnants of the initial state are visible even at $t \sim 10^6$. Inset: Correlation length $\xi$ versus time for I and II. The evolution of $\xi$ mimics panel (a). Note $\xi > 1$ at long times, indicating nontrivial quantum correlations.}
\label{fig5}
\end{figure*}

Metastability for $\lambda < 1/2$ is associated with dynamically heterogeneous relaxation, as illustrated in \reffig{fig3}(d).  The initial state is the product state of \reffig{fig2}(a), which can be written as $\rho_0 
= | \psi_{A0} \rangle \langle \psi_{A0} | \otimes 
| \psi_{B0} \rangle \langle \psi_{B0} |$ where the system is split into region $A$ containing the vacancy dimer and region $B$ containing the isolated vacancies.  The figure shows three time regimes.  Times $t_1$ are for $c(t)$ evolving from $c(0)=1$ to its plateau value.  This initial relaxation only entangles region $A$, and the state is well approximated by
$| \psi_{A}(t) \rangle \langle \psi_{A}(t) | \otimes 
| \psi_{B0} \rangle \langle \psi_{B0} |$, where 
$\vert \psi_{A}(t) \rangle = e^{-i H_{A} t} \vert \psi_{A}(0) \rangle$ with $H_A$ the restriction of \er{H} to $A$.  
Times $t_2$ correspond to the metastable regime, where region $A$ is thermalized while region $B$ is not.  The state here is $\overline{| \psi_{A}(t) \rangle \langle \psi_{A}(t) |}
 \otimes 
| \psi_{B0} \rangle \langle \psi_{B0} |
$.  Indeed, within regimes $t_1$ and $t_2$ the state $\rho(t)$ is almost entirely supported on the subspace $\mathcal{H}_{A} \otimes \vert \psi_{B0} \rangle$, where $\mathcal{H}_{A}$ indicates the Hilbert space of region $A$. Only on much longer timescales is full entanglement established between regions $A$ and $B$, see \reffig{fig3}(e). 
 
Heterogeneity in the dynamics is further confirmed by the behavior of the entanglement entropy \(S(t) = -\operatorname{Tr}\rho_A(t) \ln \rho_A(t)\), for different choices of \(A\)--\(B\) bipartition, as shown in \reffig{EntanglementEntropyPlot}. This supports the picture of propagating dimers entangling parts of the system: e.g., at \(t = 10^2\), entanglement is large for partitions that allow the dimer to visit both regions (dashed line in the left panel, and \(i = 10\) in the right), but much smaller for those where the dimer is hindered from crossing the boundary (solid line and \(i = 3\)).

%%%%%%%%%%%%%%%%%%%%%%%%%%
\noindent
{\bf \em Square-lattice quantum dimer model.}---The Hilbert space of the quantum dimer model (QDM) consists of all close-packed dimer configurations, where each site of the lattice forms a dimer with one of its nearest neighbors \cite{Rokhsar1988,Moessner2011,Chalker2017}. 
ETH in the square- and triangular-lattice QDM has recently been studied in \cite{Lan2017}. On the square lattice, the Hamiltonian is
\begin{equation}
\newcommand{\hdimers}{\includegraphics[height=2ex]{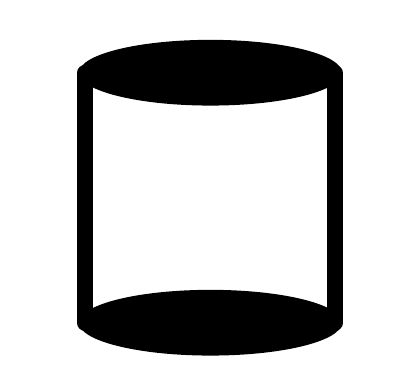}}
\newcommand{\vdimers}{\includegraphics[height=2ex]{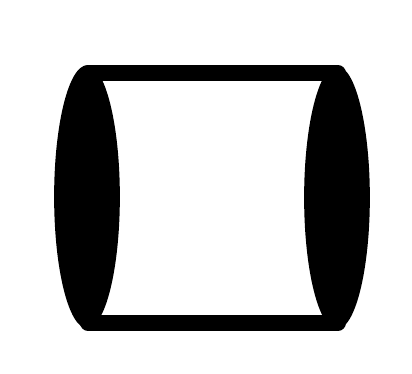}}
H\sub{QDM}= \sum_{p}\left[-(\ket{\hdimers}\bra{\vdimers} +\text{h.c.}) +
 V(\ket{\hdimers}\bra{\hdimers}+\ket{\vdimers}\bra{\vdimers})\right]\punc,
\nonumber
\end{equation}
where the sum is over all plaquettes (squares) $p$ of the lattice. The first (kinetic) term flips adjacent parallel dimers while the second (potential) counts the number of flippable plaquettes. $H\sub{QDM}$ has an RK point at $V=1$ \cite{Rokhsar1988}. A quantity conserved by \(H\sub{QDM}\) \cite{Chalker2017} -- cf.,\ the occupation \(N\) for the lattice gas -- is the flux \(\Phiv\), defined on an \(L_x\times L_y\) lattice by \(\Phi_\mu = \frac{1}{L_\mu}\sum_{\rv}(-1)^{r_x+r_y} d_{\rv\mu}\), where \(d_{\rv\mu}\) is the number of dimers, \(0\) or \(1\), on the link from site \(\rv\) in direction \(\mu = x,y\).

We consider dynamics starting from a dimer configuration and define the two-time correlation $c(t)=\sum_{\rv\mu}\langle  d_{\rv\mu}(t) d_{\rv\mu}(0) \rangle$, where the sum is over all links and the Heisenberg picture is again used. As for the lattice gas, we denote by $\overline{c(t)}$ and $[\overline{c(t)}]$ the time-integrated average and infinite-temperature average of \(c(t)\), respectively, normalized so that $\overline{c(0)}=1$ and $\overline{c(\infty)}=0$.

Figure \ref{fig5}(a) shows $\overline{c(t)}$ for all starting configurations with $\Phiv=(1,1)$ on a \(6\times 6\) lattice with periodic boundary conditions. For \(V = 0.5\), the decay of $\overline{c(t)}$ is consistently fast, while for \(V = 10\), relaxation is instead either fast or slow depending on initial configuration. The infinite-temperature average $[\overline{c(t)}]$ displays a plateau before the correlation decays to its long-time limit; \reffig{fig5}(b) shows that this plateau appears for $V\gtrsim 5$. The distinction between fast (small \(V\)) and slow (large \(V\)) dynamics is clearly visible in the lower inset of \reffig{fig5}(b), which shows the time \(\tau\) at which $[\overline{c(t)}] = \epsilon$ for \(\epsilon = 0.1,0.2,0.3\) that are below the level of the plateau (\(\simeq 0.34\)). For very large \(V\), \(\tau\) follows a power law, but with the exponent depending on \(\epsilon\). While the exponent may depend on the details of the relaxation, which involves passing through multiple steps, the presence of a power law is likely physical. The same fast--slow distinction is evident even before the appearance of the plateau, as the upper inset of \reffig{fig5}(b) shows, with a step change in the time taken to reach thresholds \(\epsilon=0.5,0.7,0.9\) that are above the plateau.

These results can be understood through a physical picture similar to that for the lattice gas, in which spatial inhomogeneities play an important role. Figure \ref{fig5}(c) shows the expectation value of the potential energy for each plaquette as time evolves, for two different initial configurations at \(V = 10\). For configuration I (top), correlations decay fast and relaxation becomes homogeneous quickly, while for the slower configuration II (bottom), heterogeneity persists even at late times. 

The inset to \reffig{fig5}(c) shows the correlation length $\xi^{2}(t) = \sum_{\rv,\rv'} {\mathcal D}^{2}(\rv\!-\!\rv') G_{\rm c}^{2}(\rv,\rv') / \sum_{\rv,\rv'} G_{\rm c}^{2}(\rv,\rv')$ where 
$G_{\rm c}(\rv,\rv') \equiv \sum_{\mu} [ \langle  d_{\rv\mu}(t) d_{\rv'\mu}(t) \rangle
- \langle  d_{\rv\mu}(t) \rangle \langle  d_{\rv'\mu}(t) \rangle ]$, and ${\mathcal D}^{2}(\rv) = L^{2} \pi^{-2} \sum_{j} \sin^{2} (\pi \rv_{j} L^{-1})$ is the lattice distance accounting for periodic boundary conditions. $\xi(t)$ eventually becomes larger than the lattice spacing, implying that neighboring degrees of freedom are correlated (unlike in the ground state at this value of $V$). The time at which $\xi(t)$ grows towards its asymptotic value coincides with the relaxation time of autocorrelators, cf.~\reffig{fig5}(a).

%%%%%%%%%%%%%%%%%%%%%%%%%%
\noindent
{\bf \em Conclusions.}--- We have demonstrated slow relaxation due to dynamical constraints in closed quantum systems without quenched disorder. 
The two models studied exhibit thermalization asymptotically, but for certain parameter values the relaxation is anomalously slow, strongly sensitive to initial conditions, and spatially heterogeneous. 
Our work should be contrasted with studies of two-component systems \cite{Yao2016,Yarloo2018}, where 
timescale separation is due to the distinction between heavy and light components. 
As in the case of classical glasses \cite{Chandler2010}, constrained dynamics -- either explicit or effective \cite{Garrahan2000,Berges2004,Chamon2005,Garrahan2009b,Castelnovo2012,Nandkishore2018} -- should be a generic mechanism for slow and spatially fluctuating relaxation in quantum systems.

%%%%%%%%%%%%%%%%%%%%%%%%%%

We thank E. Levi and M. Rigol for discussions.  This work was supported by EPSRC Grants No.\ EP/L50502X/1 (M.V.H.), No. EP/M014266/1 (J.P.G.) and No. EP/M019691/1 (Z.L. and S.P.).  

\bibliography{qconstraints}

%merlin.mbs apsrev4-1.bst 2010-07-25 4.21a (PWD, AO, DPC) hacked
%Control: key (0)
%Control: author (8) initials jnrlst
%Control: editor formatted (1) identically to author
%Control: production of article title (-1) disabled
%Control: page (0) single
%Control: year (1) truncated
%Control: production of eprint (0) enabled
\begin{thebibliography}{80}%
\makeatletter
\providecommand \@ifxundefined [1]{%
 \@ifx{#1\undefined}
}%
\providecommand \@ifnum [1]{%
 \ifnum #1\expandafter \@firstoftwo
 \else \expandafter \@secondoftwo
 \fi
}%
\providecommand \@ifx [1]{%
 \ifx #1\expandafter \@firstoftwo
 \else \expandafter \@secondoftwo
 \fi
}%
\providecommand \natexlab [1]{#1}%
\providecommand \enquote  [1]{``#1''}%
\providecommand \bibnamefont  [1]{#1}%
\providecommand \bibfnamefont [1]{#1}%
\providecommand \citenamefont [1]{#1}%
\providecommand \href@noop [0]{\@secondoftwo}%
\providecommand \href [0]{\begingroup \@sanitize@url \@href}%
\providecommand \@href[1]{\@@startlink{#1}\@@href}%
\providecommand \@@href[1]{\endgroup#1\@@endlink}%
\providecommand \@sanitize@url [0]{\catcode `\\12\catcode `\$12\catcode
  `\&12\catcode `\#12\catcode `\^12\catcode `\_12\catcode `\%12\relax}%
\providecommand \@@startlink[1]{}%
\providecommand \@@endlink[0]{}%
\providecommand \url  [0]{\begingroup\@sanitize@url \@url }%
\providecommand \@url [1]{\endgroup\@href {#1}{\urlprefix }}%
\providecommand \urlprefix  [0]{URL }%
\providecommand \Eprint [0]{\href }%
\providecommand \doibase [0]{http://dx.doi.org/}%
\providecommand \selectlanguage [0]{\@gobble}%
\providecommand \bibinfo  [0]{\@secondoftwo}%
\providecommand \bibfield  [0]{\@secondoftwo}%
\providecommand \translation [1]{[#1]}%
\providecommand \BibitemOpen [0]{}%
\providecommand \bibitemStop [0]{}%
\providecommand \bibitemNoStop [0]{.\EOS\space}%
\providecommand \EOS [0]{\spacefactor3000\relax}%
\providecommand \BibitemShut  [1]{\csname bibitem#1\endcsname}%
\let\auto@bib@innerbib\@empty
%</preamble>
\bibitem [{\citenamefont {Reimann}(2008)}]{Reimann2008}%
  \BibitemOpen
  \bibfield  {author} {\bibinfo {author} {\bibfnamefont {P.}~\bibnamefont
  {Reimann}},\ }\href {\doibase 10.1103/PhysRevLett.101.190403} {\bibfield
  {journal} {\bibinfo  {journal} {Phys. Rev. Lett.}\ }\textbf {\bibinfo
  {volume} {101}},\ \bibinfo {pages} {190403} (\bibinfo {year}
  {2008})}\BibitemShut {NoStop}%
\bibitem [{\citenamefont {Linden}\ \emph {et~al.}(2009)\citenamefont {Linden},
  \citenamefont {Popescu}, \citenamefont {Short},\ and\ \citenamefont
  {Winter}}]{Linden2009}%
  \BibitemOpen
  \bibfield  {author} {\bibinfo {author} {\bibfnamefont {N.}~\bibnamefont
  {Linden}}, \bibinfo {author} {\bibfnamefont {S.}~\bibnamefont {Popescu}},
  \bibinfo {author} {\bibfnamefont {A.~J.}\ \bibnamefont {Short}}, \ and\
  \bibinfo {author} {\bibfnamefont {A.}~\bibnamefont {Winter}},\ }\href
  {\doibase 10.1103/PhysRevE.79.061103} {\bibfield  {journal} {\bibinfo
  {journal} {Phys. Rev. E}\ }\textbf {\bibinfo {volume} {79}},\ \bibinfo
  {pages} {061103} (\bibinfo {year} {2009})}\BibitemShut {NoStop}%
\bibitem [{\citenamefont {Short}(2011)}]{Short2011}%
  \BibitemOpen
  \bibfield  {author} {\bibinfo {author} {\bibfnamefont {A.~J.}\ \bibnamefont
  {Short}},\ }\href {http://stacks.iop.org/1367-2630/13/i=5/a=053009}
  {\bibfield  {journal} {\bibinfo  {journal} {New Journal of Physics}\ }\textbf
  {\bibinfo {volume} {13}},\ \bibinfo {pages} {053009} (\bibinfo {year}
  {2011})}\BibitemShut {NoStop}%
\bibitem [{\citenamefont {Short}\ and\ \citenamefont
  {Farrelly}(2012)}]{Short2012}%
  \BibitemOpen
  \bibfield  {author} {\bibinfo {author} {\bibfnamefont {A.~J.}\ \bibnamefont
  {Short}}\ and\ \bibinfo {author} {\bibfnamefont {T.~C.}\ \bibnamefont
  {Farrelly}},\ }\href {http://stacks.iop.org/1367-2630/14/i=1/a=013063}
  {\bibfield  {journal} {\bibinfo  {journal} {New Journal of Physics}\ }\textbf
  {\bibinfo {volume} {14}},\ \bibinfo {pages} {013063} (\bibinfo {year}
  {2012})}\BibitemShut {NoStop}%
\bibitem [{\citenamefont {Reimann}\ and\ \citenamefont
  {Kastner}(2012)}]{Reimann2012}%
  \BibitemOpen
  \bibfield  {author} {\bibinfo {author} {\bibfnamefont {P.}~\bibnamefont
  {Reimann}}\ and\ \bibinfo {author} {\bibfnamefont {M.}~\bibnamefont
  {Kastner}},\ }\href {http://stacks.iop.org/1367-2630/14/i=4/a=043020}
  {\bibfield  {journal} {\bibinfo  {journal} {New Journal of Physics}\ }\textbf
  {\bibinfo {volume} {14}},\ \bibinfo {pages} {043020} (\bibinfo {year}
  {2012})}\BibitemShut {NoStop}%
\bibitem [{\citenamefont {Gogolin}\ and\ \citenamefont
  {Eisert}(2016)}]{Gogolin2016}%
  \BibitemOpen
  \bibfield  {author} {\bibinfo {author} {\bibfnamefont {C.}~\bibnamefont
  {Gogolin}}\ and\ \bibinfo {author} {\bibfnamefont {J.}~\bibnamefont
  {Eisert}},\ }\href {http://stacks.iop.org/0034-4885/79/i=5/a=056001}
  {\bibfield  {journal} {\bibinfo  {journal} {Reports on Progress in Physics}\
  }\textbf {\bibinfo {volume} {79}},\ \bibinfo {pages} {056001} (\bibinfo
  {year} {2016})}\BibitemShut {NoStop}%
\bibitem [{\citenamefont {D'Alessio}\ \emph {et~al.}(2016)\citenamefont
  {D'Alessio}, \citenamefont {Kafri}, \citenamefont {Polkovnikov},\ and\
  \citenamefont {Rigol}}]{DAlessio2016}%
  \BibitemOpen
  \bibfield  {author} {\bibinfo {author} {\bibfnamefont {L.}~\bibnamefont
  {D'Alessio}}, \bibinfo {author} {\bibfnamefont {Y.}~\bibnamefont {Kafri}},
  \bibinfo {author} {\bibfnamefont {A.}~\bibnamefont {Polkovnikov}}, \ and\
  \bibinfo {author} {\bibfnamefont {M.}~\bibnamefont {Rigol}},\ }\href@noop {}
  {\bibfield  {journal} {\bibinfo  {journal} {Adv. Phys.}\ }\textbf {\bibinfo
  {volume} {65}},\ \bibinfo {pages} {239} (\bibinfo {year} {2016})}\BibitemShut
  {NoStop}%
\bibitem [{\citenamefont {Borgonovi}\ \emph {et~al.}(2016)\citenamefont
  {Borgonovi}, \citenamefont {Izrailev}, \citenamefont {Santos},\ and\
  \citenamefont {Zelevinsky}}]{Borgonovi2016}%
  \BibitemOpen
  \bibfield  {author} {\bibinfo {author} {\bibfnamefont {F.}~\bibnamefont
  {Borgonovi}}, \bibinfo {author} {\bibfnamefont {F.~M.}\ \bibnamefont
  {Izrailev}}, \bibinfo {author} {\bibfnamefont {L.~F.}\ \bibnamefont
  {Santos}}, \ and\ \bibinfo {author} {\bibfnamefont {V.~G.}\ \bibnamefont
  {Zelevinsky}},\ }\bibfield  {booktitle} {\emph {\bibinfo {booktitle} {Quantum
  chaos and thermalization in isolated systems of interacting particles}},\
  }\href {\doibase https://doi.org/10.1016/j.physrep.2016.02.005} {\bibfield
  {journal} {\bibinfo  {journal} {Physics Reports}\ }\textbf {\bibinfo {volume}
  {626}},\ \bibinfo {pages} {1} (\bibinfo {year} {2016})}\BibitemShut {NoStop}%
\bibitem [{\citenamefont {Deutsch}(1991)}]{Deutsch1991}%
  \BibitemOpen
  \bibfield  {author} {\bibinfo {author} {\bibfnamefont {J.~M.}\ \bibnamefont
  {Deutsch}},\ }\href {\doibase 10.1103/PhysRevA.43.2046} {\bibfield  {journal}
  {\bibinfo  {journal} {Phys. Rev. A}\ }\textbf {\bibinfo {volume} {43}},\
  \bibinfo {pages} {2046} (\bibinfo {year} {1991})}\BibitemShut {NoStop}%
\bibitem [{\citenamefont {Srednicki}(1994)}]{Srednicki1994}%
  \BibitemOpen
  \bibfield  {author} {\bibinfo {author} {\bibfnamefont {M.}~\bibnamefont
  {Srednicki}},\ }\href {\doibase 10.1103/PhysRevE.50.888} {\bibfield
  {journal} {\bibinfo  {journal} {Phys. Rev. E}\ }\textbf {\bibinfo {volume}
  {50}},\ \bibinfo {pages} {888} (\bibinfo {year} {1994})}\BibitemShut
  {NoStop}%
\bibitem [{\citenamefont {Tasaki}(1998)}]{Tasaki1998}%
  \BibitemOpen
  \bibfield  {author} {\bibinfo {author} {\bibfnamefont {H.}~\bibnamefont
  {Tasaki}},\ }\href {\doibase 10.1103/PhysRevLett.80.1373} {\bibfield
  {journal} {\bibinfo  {journal} {Phys. Rev. Lett.}\ }\textbf {\bibinfo
  {volume} {80}},\ \bibinfo {pages} {1373} (\bibinfo {year}
  {1998})}\BibitemShut {NoStop}%
\bibitem [{\citenamefont {Rigol}\ \emph {et~al.}(2008)\citenamefont {Rigol},
  \citenamefont {Dunjko},\ and\ \citenamefont {Olshanii}}]{Rigol2008}%
  \BibitemOpen
  \bibfield  {author} {\bibinfo {author} {\bibfnamefont {M.}~\bibnamefont
  {Rigol}}, \bibinfo {author} {\bibfnamefont {V.}~\bibnamefont {Dunjko}}, \
  and\ \bibinfo {author} {\bibfnamefont {M.}~\bibnamefont {Olshanii}},\ }\href
  {\doibase 10.1038/nature06838} {\bibfield  {journal} {\bibinfo  {journal}
  {Nature}\ }\textbf {\bibinfo {volume} {452}},\ \bibinfo {pages} {854}
  (\bibinfo {year} {2008})}\BibitemShut {NoStop}%
\bibitem [{\citenamefont {Essler}\ and\ \citenamefont
  {Fagotti}(2016)}]{Essler2016}%
  \BibitemOpen
  \bibfield  {author} {\bibinfo {author} {\bibfnamefont {F.~H.~L.}\
  \bibnamefont {Essler}}\ and\ \bibinfo {author} {\bibfnamefont
  {M.}~\bibnamefont {Fagotti}},\ }\href@noop {} {\bibfield  {journal} {\bibinfo
   {journal} {J. Stat. Mech.}\ }\textbf {\bibinfo {volume} {2016}},\ \bibinfo
  {pages} {064002} (\bibinfo {year} {2016})}\BibitemShut {NoStop}%
\bibitem [{\citenamefont {Rigol}\ \emph {et~al.}(2007)\citenamefont {Rigol},
  \citenamefont {Dunjko}, \citenamefont {Yurovsky},\ and\ \citenamefont
  {Olshanii}}]{Rigol2007}%
  \BibitemOpen
  \bibfield  {author} {\bibinfo {author} {\bibfnamefont {M.}~\bibnamefont
  {Rigol}}, \bibinfo {author} {\bibfnamefont {V.}~\bibnamefont {Dunjko}},
  \bibinfo {author} {\bibfnamefont {V.}~\bibnamefont {Yurovsky}}, \ and\
  \bibinfo {author} {\bibfnamefont {M.}~\bibnamefont {Olshanii}},\ }\href
  {\doibase 10.1103/PhysRevLett.98.050405} {\bibfield  {journal} {\bibinfo
  {journal} {Phys. Rev. Lett.}\ }\textbf {\bibinfo {volume} {98}},\ \bibinfo
  {pages} {050405} (\bibinfo {year} {2007})}\BibitemShut {NoStop}%
\bibitem [{\citenamefont {Vidmar}\ and\ \citenamefont
  {Rigol}(2016)}]{Vidmar2016}%
  \BibitemOpen
  \bibfield  {author} {\bibinfo {author} {\bibfnamefont {L.}~\bibnamefont
  {Vidmar}}\ and\ \bibinfo {author} {\bibfnamefont {M.}~\bibnamefont {Rigol}},\
  }\href@noop {} {\bibfield  {journal} {\bibinfo  {journal} {J. Stat. Mech.}\
  }\textbf {\bibinfo {volume} {2016}},\ \bibinfo {pages} {064007} (\bibinfo
  {year} {2016})}\BibitemShut {NoStop}%
\bibitem [{\citenamefont {Altshuler}\ \emph {et~al.}(1997)\citenamefont
  {Altshuler}, \citenamefont {Gefen}, \citenamefont {Kamenev},\ and\
  \citenamefont {Levitov}}]{Altshuler1997}%
  \BibitemOpen
  \bibfield  {author} {\bibinfo {author} {\bibfnamefont {B.~L.}\ \bibnamefont
  {Altshuler}}, \bibinfo {author} {\bibfnamefont {Y.}~\bibnamefont {Gefen}},
  \bibinfo {author} {\bibfnamefont {A.}~\bibnamefont {Kamenev}}, \ and\
  \bibinfo {author} {\bibfnamefont {L.~S.}\ \bibnamefont {Levitov}},\
  }\href@noop {} {\bibfield  {journal} {\bibinfo  {journal} {Phys. Rev. Lett.}\
  }\textbf {\bibinfo {volume} {78}},\ \bibinfo {pages} {2803} (\bibinfo {year}
  {1997})}\BibitemShut {NoStop}%
\bibitem [{\citenamefont {Basko}\ \emph {et~al.}(2006)\citenamefont {Basko},
  \citenamefont {Aleiner},\ and\ \citenamefont {Altshuler}}]{Basko2006}%
  \BibitemOpen
  \bibfield  {author} {\bibinfo {author} {\bibfnamefont {D.}~\bibnamefont
  {Basko}}, \bibinfo {author} {\bibfnamefont {I.}~\bibnamefont {Aleiner}}, \
  and\ \bibinfo {author} {\bibfnamefont {B.}~\bibnamefont {Altshuler}},\
  }\href@noop {} {\bibfield  {journal} {\bibinfo  {journal} {Ann. of Phys.}\
  }\textbf {\bibinfo {volume} {321}},\ \bibinfo {pages} {1126} (\bibinfo {year}
  {2006})}\BibitemShut {NoStop}%
\bibitem [{\citenamefont {Gornyi}\ \emph {et~al.}(2005)\citenamefont {Gornyi},
  \citenamefont {Mirlin},\ and\ \citenamefont {Polyakov}}]{Gornyi2005}%
  \BibitemOpen
  \bibfield  {author} {\bibinfo {author} {\bibfnamefont {I.}~\bibnamefont
  {Gornyi}}, \bibinfo {author} {\bibfnamefont {A.}~\bibnamefont {Mirlin}}, \
  and\ \bibinfo {author} {\bibfnamefont {D.}~\bibnamefont {Polyakov}},\ }\href
  {\doibase 10.1103/PhysRevLett.95.206603} {\bibfield  {journal} {\bibinfo
  {journal} {Phys. Rev. Lett.}\ }\textbf {\bibinfo {volume} {95}},\ \bibinfo
  {pages} {206603} (\bibinfo {year} {2005})}\BibitemShut {NoStop}%
\bibitem [{\citenamefont {Oganesyan}\ and\ \citenamefont
  {Huse}(2007)}]{Oganesyan2007}%
  \BibitemOpen
  \bibfield  {author} {\bibinfo {author} {\bibfnamefont {V.}~\bibnamefont
  {Oganesyan}}\ and\ \bibinfo {author} {\bibfnamefont {D.~A.}\ \bibnamefont
  {Huse}},\ }\href {\doibase 10.1103/PhysRevB.75.155111} {\bibfield  {journal}
  {\bibinfo  {journal} {Phys. Rev. B}\ }\textbf {\bibinfo {volume} {75}},\
  \bibinfo {pages} {155111} (\bibinfo {year} {2007})}\BibitemShut {NoStop}%
\bibitem [{\citenamefont {Znidaric}\ \emph {et~al.}(2008)\citenamefont
  {Znidaric}, \citenamefont {Prosen},\ and\ \citenamefont
  {Prelovsek}}]{Znidaric2008}%
  \BibitemOpen
  \bibfield  {author} {\bibinfo {author} {\bibfnamefont {M.}~\bibnamefont
  {Znidaric}}, \bibinfo {author} {\bibfnamefont {T.}~\bibnamefont {Prosen}}, \
  and\ \bibinfo {author} {\bibfnamefont {P.}~\bibnamefont {Prelovsek}},\
  }\href@noop {} {\bibfield  {journal} {\bibinfo  {journal} {Phys. Rev. B}\
  }\textbf {\bibinfo {volume} {77}},\ \bibinfo {pages} {064426} (\bibinfo
  {year} {2008})}\BibitemShut {NoStop}%
\bibitem [{\citenamefont {Pal}\ and\ \citenamefont {Huse}(2010)}]{Pal2010}%
  \BibitemOpen
  \bibfield  {author} {\bibinfo {author} {\bibfnamefont {A.}~\bibnamefont
  {Pal}}\ and\ \bibinfo {author} {\bibfnamefont {D.~A.}\ \bibnamefont {Huse}},\
  }\href@noop {} {\bibfield  {journal} {\bibinfo  {journal} {Phys. Rev. B}\
  }\textbf {\bibinfo {volume} {82}},\ \bibinfo {pages} {174411} (\bibinfo
  {year} {2010})}\BibitemShut {NoStop}%
\bibitem [{\citenamefont {Bardarson}\ \emph {et~al.}(2012)\citenamefont
  {Bardarson}, \citenamefont {Pollmann},\ and\ \citenamefont
  {Moore}}]{Bardarson2012}%
  \BibitemOpen
  \bibfield  {author} {\bibinfo {author} {\bibfnamefont {J.~H.}\ \bibnamefont
  {Bardarson}}, \bibinfo {author} {\bibfnamefont {F.}~\bibnamefont {Pollmann}},
  \ and\ \bibinfo {author} {\bibfnamefont {J.~E.}\ \bibnamefont {Moore}},\
  }\href {\doibase 10.1103/PhysRevLett.109.017202} {\bibfield  {journal}
  {\bibinfo  {journal} {Phys. Rev. Lett.}\ }\textbf {\bibinfo {volume} {109}},\
  \bibinfo {pages} {017202} (\bibinfo {year} {2012})}\BibitemShut {NoStop}%
\bibitem [{\citenamefont {Serbyn}\ \emph {et~al.}(2013)\citenamefont {Serbyn},
  \citenamefont {{Papi\ifmmode \acute{c}\else {\'c}\fi{}}},\ and\ \citenamefont
  {Abanin}}]{Serbyn2013}%
  \BibitemOpen
  \bibfield  {author} {\bibinfo {author} {\bibfnamefont {M.}~\bibnamefont
  {Serbyn}}, \bibinfo {author} {\bibfnamefont {Z.}~\bibnamefont {{Papi\ifmmode
  \acute{c}\else {\'c}\fi{}}}}, \ and\ \bibinfo {author} {\bibfnamefont
  {D.~A.}\ \bibnamefont {Abanin}},\ }\href {\doibase
  10.1103/PhysRevLett.110.260601} {\bibfield  {journal} {\bibinfo  {journal}
  {Phys. Rev. Lett.}\ }\textbf {\bibinfo {volume} {110}},\ \bibinfo {pages}
  {260601} (\bibinfo {year} {2013})}\BibitemShut {NoStop}%
\bibitem [{\citenamefont {Huse}\ \emph {et~al.}(2014)\citenamefont {Huse},
  \citenamefont {Nandkishore},\ and\ \citenamefont {Oganesyan}}]{Huse2014}%
  \BibitemOpen
  \bibfield  {author} {\bibinfo {author} {\bibfnamefont {D.~A.}\ \bibnamefont
  {Huse}}, \bibinfo {author} {\bibfnamefont {R.}~\bibnamefont {Nandkishore}}, \
  and\ \bibinfo {author} {\bibfnamefont {V.}~\bibnamefont {Oganesyan}},\ }\href
  {\doibase 10.1103/PhysRevB.90.174202} {\bibfield  {journal} {\bibinfo
  {journal} {Phys. Rev. B}\ }\textbf {\bibinfo {volume} {90}},\ \bibinfo
  {pages} {174202} (\bibinfo {year} {2014})}\BibitemShut {NoStop}%
\bibitem [{\citenamefont {Andraschko}\ \emph {et~al.}(2014)\citenamefont
  {Andraschko}, \citenamefont {Enss},\ and\ \citenamefont
  {Sirker}}]{Andraschko2014}%
  \BibitemOpen
  \bibfield  {author} {\bibinfo {author} {\bibfnamefont {F.}~\bibnamefont
  {Andraschko}}, \bibinfo {author} {\bibfnamefont {T.}~\bibnamefont {Enss}}, \
  and\ \bibinfo {author} {\bibfnamefont {J.}~\bibnamefont {Sirker}},\ }\href
  {\doibase 10.1103/PhysRevLett.113.217201} {\bibfield  {journal} {\bibinfo
  {journal} {Phys. Rev. Lett.}\ }\textbf {\bibinfo {volume} {113}},\ \bibinfo
  {pages} {217201} (\bibinfo {year} {2014})}\BibitemShut {NoStop}%
\bibitem [{\citenamefont {Yao}\ \emph {et~al.}(2014)\citenamefont {Yao},
  \citenamefont {Laumann}, \citenamefont {Gopalakrishnan}, \citenamefont
  {Knap}, \citenamefont {M{\"u}ller}, \citenamefont {Demler},\ and\
  \citenamefont {Lukin}}]{Yao2014}%
  \BibitemOpen
  \bibfield  {author} {\bibinfo {author} {\bibfnamefont {N.~Y.}\ \bibnamefont
  {Yao}}, \bibinfo {author} {\bibfnamefont {C.~R.}\ \bibnamefont {Laumann}},
  \bibinfo {author} {\bibfnamefont {S.}~\bibnamefont {Gopalakrishnan}},
  \bibinfo {author} {\bibfnamefont {M.}~\bibnamefont {Knap}}, \bibinfo {author}
  {\bibfnamefont {M.}~\bibnamefont {M{\"u}ller}}, \bibinfo {author}
  {\bibfnamefont {E.~A.}\ \bibnamefont {Demler}}, \ and\ \bibinfo {author}
  {\bibfnamefont {M.~D.}\ \bibnamefont {Lukin}},\ }\href {\doibase
  10.1103/PhysRevLett.113.243002} {\bibfield  {journal} {\bibinfo  {journal}
  {Phys. Rev. Lett.}\ }\textbf {\bibinfo {volume} {113}},\ \bibinfo {pages}
  {243002} (\bibinfo {year} {2014})}\BibitemShut {NoStop}%
\bibitem [{\citenamefont {Serbyn}\ \emph {et~al.}(2014)\citenamefont {Serbyn},
  \citenamefont {Papic},\ and\ \citenamefont {Abanin}}]{Serbyn2014}%
  \BibitemOpen
  \bibfield  {author} {\bibinfo {author} {\bibfnamefont {M.}~\bibnamefont
  {Serbyn}}, \bibinfo {author} {\bibfnamefont {Z.}~\bibnamefont {Papic}}, \
  and\ \bibinfo {author} {\bibfnamefont {D.~A.}\ \bibnamefont {Abanin}},\
  }\href {\doibase 10.1103/PhysRevB.90.174302} {\bibfield  {journal} {\bibinfo
  {journal} {Phys. Rev. B}\ }\textbf {\bibinfo {volume} {90}},\ \bibinfo
  {pages} {174302} (\bibinfo {year} {2014})}\BibitemShut {NoStop}%
\bibitem [{\citenamefont {Laumann}\ \emph {et~al.}(2014)\citenamefont
  {Laumann}, \citenamefont {Pal},\ and\ \citenamefont
  {Scardicchio}}]{Laumann2014}%
  \BibitemOpen
  \bibfield  {author} {\bibinfo {author} {\bibfnamefont {C.~R.}\ \bibnamefont
  {Laumann}}, \bibinfo {author} {\bibfnamefont {A.}~\bibnamefont {Pal}}, \ and\
  \bibinfo {author} {\bibfnamefont {A.}~\bibnamefont {Scardicchio}},\ }\href
  {\doibase 10.1103/PhysRevLett.113.200405} {\bibfield  {journal} {\bibinfo
  {journal} {Phys. Rev. Lett.}\ }\textbf {\bibinfo {volume} {113}},\ \bibinfo
  {pages} {200405} (\bibinfo {year} {2014})}\BibitemShut {NoStop}%
\bibitem [{\citenamefont {{De Roeck}}\ and\ \citenamefont
  {Huveneers}(2014)}]{De-Roeck2014}%
  \BibitemOpen
  \bibfield  {author} {\bibinfo {author} {\bibfnamefont {W.}~\bibnamefont {{De
  Roeck}}}\ and\ \bibinfo {author} {\bibfnamefont {F.}~\bibnamefont
  {Huveneers}},\ }\href {\doibase 10.1103/PhysRevB.90.165137} {\bibfield
  {journal} {\bibinfo  {journal} {Phys. Rev. B}\ }\textbf {\bibinfo {volume}
  {90}},\ \bibinfo {pages} {165137} (\bibinfo {year} {2014})}\BibitemShut
  {NoStop}%
\bibitem [{\citenamefont {Ros}\ \emph {et~al.}(2015)\citenamefont {Ros},
  \citenamefont {M{\"u}ller},\ and\ \citenamefont
  {Scardicchio}}]{Scardicchio2015}%
  \BibitemOpen
  \bibfield  {author} {\bibinfo {author} {\bibfnamefont {V.}~\bibnamefont
  {Ros}}, \bibinfo {author} {\bibfnamefont {M.}~\bibnamefont {M{\"u}ller}}, \
  and\ \bibinfo {author} {\bibfnamefont {A.}~\bibnamefont {Scardicchio}},\
  }\href {\doibase http://dx.doi.org/10.1016/j.nuclphysb.2014.12.014}
  {\bibfield  {journal} {\bibinfo  {journal} {Nuclear Physics B}\ }\textbf
  {\bibinfo {volume} {891}},\ \bibinfo {pages} {420 } (\bibinfo {year}
  {2015})}\BibitemShut {NoStop}%
\bibitem [{\citenamefont {Vasseur}\ \emph {et~al.}(2015)\citenamefont
  {Vasseur}, \citenamefont {Parameswaran},\ and\ \citenamefont
  {Moore}}]{Vasseur2015}%
  \BibitemOpen
  \bibfield  {author} {\bibinfo {author} {\bibfnamefont {R.}~\bibnamefont
  {Vasseur}}, \bibinfo {author} {\bibfnamefont {S.~A.}\ \bibnamefont
  {Parameswaran}}, \ and\ \bibinfo {author} {\bibfnamefont {J.~E.}\
  \bibnamefont {Moore}},\ }\href {\doibase 10.1103/PhysRevB.91.140202}
  {\bibfield  {journal} {\bibinfo  {journal} {Phys. Rev. B}\ }\textbf {\bibinfo
  {volume} {91}},\ \bibinfo {pages} {140202} (\bibinfo {year}
  {2015})}\BibitemShut {NoStop}%
\bibitem [{\citenamefont {Agarwal}\ \emph {et~al.}(2015)\citenamefont
  {Agarwal}, \citenamefont {Gopalakrishnan}, \citenamefont {Knap},
  \citenamefont {M{\"u}ller},\ and\ \citenamefont {Demler}}]{Agarwal2015}%
  \BibitemOpen
  \bibfield  {author} {\bibinfo {author} {\bibfnamefont {K.}~\bibnamefont
  {Agarwal}}, \bibinfo {author} {\bibfnamefont {S.}~\bibnamefont
  {Gopalakrishnan}}, \bibinfo {author} {\bibfnamefont {M.}~\bibnamefont
  {Knap}}, \bibinfo {author} {\bibfnamefont {M.}~\bibnamefont {M{\"u}ller}}, \
  and\ \bibinfo {author} {\bibfnamefont {E.}~\bibnamefont {Demler}},\ }\href
  {\doibase 10.1103/PhysRevLett.114.160401} {\bibfield  {journal} {\bibinfo
  {journal} {Phys. Rev. Lett.}\ }\textbf {\bibinfo {volume} {114}},\ \bibinfo
  {pages} {160401} (\bibinfo {year} {2015})}\BibitemShut {NoStop}%
\bibitem [{\citenamefont {{Bar Lev}}\ \emph {et~al.}(2015)\citenamefont {{Bar
  Lev}}, \citenamefont {Cohen},\ and\ \citenamefont {Reichman}}]{Bar-Lev2015}%
  \BibitemOpen
  \bibfield  {author} {\bibinfo {author} {\bibfnamefont {Y.}~\bibnamefont {{Bar
  Lev}}}, \bibinfo {author} {\bibfnamefont {G.}~\bibnamefont {Cohen}}, \ and\
  \bibinfo {author} {\bibfnamefont {D.~R.}\ \bibnamefont {Reichman}},\ }\href
  {\doibase 10.1103/PhysRevLett.114.100601} {\bibfield  {journal} {\bibinfo
  {journal} {Phys. Rev. Lett.}\ }\textbf {\bibinfo {volume} {114}},\ \bibinfo
  {pages} {100601} (\bibinfo {year} {2015})}\BibitemShut {NoStop}%
\bibitem [{\citenamefont {Imbrie}(2016)}]{Imbrie2016}%
  \BibitemOpen
  \bibfield  {author} {\bibinfo {author} {\bibfnamefont {J.~Z.}\ \bibnamefont
  {Imbrie}},\ }\href@noop {} {\bibfield  {journal} {\bibinfo  {journal} {J.
  Stat. Phys.}\ }\textbf {\bibinfo {volume} {163}},\ \bibinfo {pages} {998}
  (\bibinfo {year} {2016})}\BibitemShut {NoStop}%
\bibitem [{\citenamefont {Schreiber}\ \emph {et~al.}(2015)\citenamefont
  {Schreiber}, \citenamefont {Hodgman}, \citenamefont {Bordia}, \citenamefont
  {L{\~A}¼schen}, \citenamefont {Fischer}, \citenamefont {Vosk}, \citenamefont
  {Altman}, \citenamefont {Schneider},\ and\ \citenamefont
  {Bloch}}]{Schreiber2015}%
  \BibitemOpen
  \bibfield  {author} {\bibinfo {author} {\bibfnamefont {M.}~\bibnamefont
  {Schreiber}}, \bibinfo {author} {\bibfnamefont {S.~S.}\ \bibnamefont
  {Hodgman}}, \bibinfo {author} {\bibfnamefont {P.}~\bibnamefont {Bordia}},
  \bibinfo {author} {\bibfnamefont {H.~P.}\ \bibnamefont {L{\~A}¼schen}},
  \bibinfo {author} {\bibfnamefont {M.~H.}\ \bibnamefont {Fischer}}, \bibinfo
  {author} {\bibfnamefont {R.}~\bibnamefont {Vosk}}, \bibinfo {author}
  {\bibfnamefont {E.}~\bibnamefont {Altman}}, \bibinfo {author} {\bibfnamefont
  {U.}~\bibnamefont {Schneider}}, \ and\ \bibinfo {author} {\bibfnamefont
  {I.}~\bibnamefont {Bloch}},\ }\href@noop {} {\bibfield  {journal} {\bibinfo
  {journal} {Science}\ }\textbf {\bibinfo {volume} {349}},\ \bibinfo {pages}
  {842} (\bibinfo {year} {2015})}\BibitemShut {NoStop}%
\bibitem [{\citenamefont {Bordia}\ \emph {et~al.}(2016)\citenamefont {Bordia},
  \citenamefont {Luschen}, \citenamefont {Hodgman}, \citenamefont {Schreiber},
  \citenamefont {Bloch},\ and\ \citenamefont {Schneider}}]{Bordia2016}%
  \BibitemOpen
  \bibfield  {author} {\bibinfo {author} {\bibfnamefont {P.}~\bibnamefont
  {Bordia}}, \bibinfo {author} {\bibfnamefont {H.~P.}\ \bibnamefont {Luschen}},
  \bibinfo {author} {\bibfnamefont {S.~S.}\ \bibnamefont {Hodgman}}, \bibinfo
  {author} {\bibfnamefont {M.}~\bibnamefont {Schreiber}}, \bibinfo {author}
  {\bibfnamefont {I.}~\bibnamefont {Bloch}}, \ and\ \bibinfo {author}
  {\bibfnamefont {U.}~\bibnamefont {Schneider}},\ }\href {\doibase
  10.1103/PhysRevLett.116.140401} {\bibfield  {journal} {\bibinfo  {journal}
  {Phys. Rev. Lett.}\ }\textbf {\bibinfo {volume} {116}},\ \bibinfo {pages}
  {140401} (\bibinfo {year} {2016})}\BibitemShut {NoStop}%
\bibitem [{\citenamefont {Smith}\ \emph {et~al.}(2016)\citenamefont {Smith},
  \citenamefont {Lee}, \citenamefont {Richerme}, \citenamefont {Neyenhuis},
  \citenamefont {Hess}, \citenamefont {Hauke}, \citenamefont {Heyl},
  \citenamefont {Huse},\ and\ \citenamefont {Monroe}}]{Smith2016}%
  \BibitemOpen
  \bibfield  {author} {\bibinfo {author} {\bibfnamefont {J.}~\bibnamefont
  {Smith}}, \bibinfo {author} {\bibfnamefont {A.}~\bibnamefont {Lee}}, \bibinfo
  {author} {\bibfnamefont {P.}~\bibnamefont {Richerme}}, \bibinfo {author}
  {\bibfnamefont {B.}~\bibnamefont {Neyenhuis}}, \bibinfo {author}
  {\bibfnamefont {P.~W.}\ \bibnamefont {Hess}}, \bibinfo {author}
  {\bibfnamefont {P.}~\bibnamefont {Hauke}}, \bibinfo {author} {\bibfnamefont
  {M.}~\bibnamefont {Heyl}}, \bibinfo {author} {\bibfnamefont {D.~A.}\
  \bibnamefont {Huse}}, \ and\ \bibinfo {author} {\bibfnamefont
  {C.}~\bibnamefont {Monroe}},\ }\href@noop {} {\bibfield  {journal} {\bibinfo
  {journal} {Nature Phys.}\ }\textbf {\bibinfo {volume} {12}},\ \bibinfo
  {pages} {907} (\bibinfo {year} {2016})}\BibitemShut {NoStop}%
\bibitem [{\citenamefont {Nandkishore}\ and\ \citenamefont
  {Huse}(2015)}]{Nandkishore2015}%
  \BibitemOpen
  \bibfield  {author} {\bibinfo {author} {\bibfnamefont {R.}~\bibnamefont
  {Nandkishore}}\ and\ \bibinfo {author} {\bibfnamefont {D.~A.}\ \bibnamefont
  {Huse}},\ }\href {\doibase 10.1146/annurev-conmatphys-031214-014726}
  {\bibfield  {journal} {\bibinfo  {journal} {Annu. Rev. Condens. Matter
  Phys.}\ }\textbf {\bibinfo {volume} {6}},\ \bibinfo {pages} {15} (\bibinfo
  {year} {2015})}\BibitemShut {NoStop}%
\bibitem [{\citenamefont {Altman}\ and\ \citenamefont
  {Vosk}(2015)}]{Altman2015}%
  \BibitemOpen
  \bibfield  {author} {\bibinfo {author} {\bibfnamefont {E.}~\bibnamefont
  {Altman}}\ and\ \bibinfo {author} {\bibfnamefont {R.}~\bibnamefont {Vosk}},\
  }\href@noop {} {\bibfield  {journal} {\bibinfo  {journal} {Annu. Rev.
  Condens. Matter Phys.}\ }\textbf {\bibinfo {volume} {6}},\ \bibinfo {pages}
  {383} (\bibinfo {year} {2015})}\BibitemShut {NoStop}%
\bibitem [{\citenamefont {Abanin}\ and\ \citenamefont
  {Papic}(2017)}]{Abanin2017}%
  \BibitemOpen
  \bibfield  {author} {\bibinfo {author} {\bibfnamefont {D.~A.}\ \bibnamefont
  {Abanin}}\ and\ \bibinfo {author} {\bibfnamefont {Z.}~\bibnamefont {Papic}},\
  }\href@noop {} {\bibfield  {journal} {\bibinfo  {journal} {Ann. Phys.
  (Berlin)}\ }\textbf {\bibinfo {volume} {529}},\ \bibinfo {pages} {1700169}
  (\bibinfo {year} {2017})}\BibitemShut {NoStop}%
\bibitem [{\citenamefont {Binder}\ and\ \citenamefont
  {Young}(1986)}]{Binder1986}%
  \BibitemOpen
  \bibfield  {author} {\bibinfo {author} {\bibfnamefont {K.}~\bibnamefont
  {Binder}}\ and\ \bibinfo {author} {\bibfnamefont {A.~P.}\ \bibnamefont
  {Young}},\ }\href@noop {} {\bibfield  {journal} {\bibinfo  {journal} {Rev.
  Mod. Phys.}\ }\textbf {\bibinfo {volume} {58}},\ \bibinfo {pages} {801}
  (\bibinfo {year} {1986})}\BibitemShut {NoStop}%
\bibitem [{\citenamefont {Binder}\ and\ \citenamefont
  {Kob}(2011)}]{Binder2011}%
  \BibitemOpen
  \bibfield  {author} {\bibinfo {author} {\bibfnamefont {K.}~\bibnamefont
  {Binder}}\ and\ \bibinfo {author} {\bibfnamefont {W.}~\bibnamefont {Kob}},\
  }\href@noop {} {\emph {\bibinfo {title} {{Glassy Materials and Disordered
  Solids}}}}\ (\bibinfo  {publisher} {World Scientific},\ \bibinfo {year}
  {2011})\BibitemShut {NoStop}%
\bibitem [{\citenamefont {Berthier}\ and\ \citenamefont
  {Biroli}(2011)}]{Berthier2011}%
  \BibitemOpen
  \bibfield  {author} {\bibinfo {author} {\bibfnamefont {L.}~\bibnamefont
  {Berthier}}\ and\ \bibinfo {author} {\bibfnamefont {G.}~\bibnamefont
  {Biroli}},\ }\href {\doibase 10.1103/RevModPhys.83.587} {\bibfield  {journal}
  {\bibinfo  {journal} {Rev. Mod. Phys.}\ }\textbf {\bibinfo {volume} {83}},\
  \bibinfo {pages} {587} (\bibinfo {year} {2011})}\BibitemShut {NoStop}%
\bibitem [{\citenamefont {Biroli}\ and\ \citenamefont
  {Garrahan}(2013)}]{Biroli2013}%
  \BibitemOpen
  \bibfield  {author} {\bibinfo {author} {\bibfnamefont {G.}~\bibnamefont
  {Biroli}}\ and\ \bibinfo {author} {\bibfnamefont {J.~P.}\ \bibnamefont
  {Garrahan}},\ }\href {\doibase 10.1063/1.4795539} {\bibfield  {journal}
  {\bibinfo  {journal} {J. Chem. Phys.}\ }\textbf {\bibinfo {volume} {138}},\
  \bibinfo {eid} {12A301} (\bibinfo {year} {2013})}\BibitemShut {NoStop}%
\bibitem [{\citenamefont {Fredrickson}\ and\ \citenamefont
  {Andersen}(1984)}]{Fredrickson1984}%
  \BibitemOpen
  \bibfield  {author} {\bibinfo {author} {\bibfnamefont {G.~H.}\ \bibnamefont
  {Fredrickson}}\ and\ \bibinfo {author} {\bibfnamefont {H.~C.}\ \bibnamefont
  {Andersen}},\ }\href@noop {} {\bibfield  {journal} {\bibinfo  {journal}
  {Phys. Rev. Lett.}\ }\textbf {\bibinfo {volume} {53}},\ \bibinfo {pages}
  {1244} (\bibinfo {year} {1984})}\BibitemShut {NoStop}%
\bibitem [{\citenamefont {Palmer}\ \emph {et~al.}(1984)\citenamefont {Palmer},
  \citenamefont {Stein}, \citenamefont {Abrahams},\ and\ \citenamefont
  {Anderson}}]{Palmer1984}%
  \BibitemOpen
  \bibfield  {author} {\bibinfo {author} {\bibfnamefont {R.~G.}\ \bibnamefont
  {Palmer}}, \bibinfo {author} {\bibfnamefont {D.~L.}\ \bibnamefont {Stein}},
  \bibinfo {author} {\bibfnamefont {E.}~\bibnamefont {Abrahams}}, \ and\
  \bibinfo {author} {\bibfnamefont {P.~W.}\ \bibnamefont {Anderson}},\
  }\href@noop {} {\bibfield  {journal} {\bibinfo  {journal} {Phys. Rev. Lett.}\
  }\textbf {\bibinfo {volume} {53}},\ \bibinfo {pages} {958} (\bibinfo {year}
  {1984})}\BibitemShut {NoStop}%
\bibitem [{\citenamefont {Ritort}\ and\ \citenamefont
  {Sollich}(2003)}]{Ritort2003}%
  \BibitemOpen
  \bibfield  {author} {\bibinfo {author} {\bibfnamefont {F.}~\bibnamefont
  {Ritort}}\ and\ \bibinfo {author} {\bibfnamefont {P.}~\bibnamefont
  {Sollich}},\ }\href {\doibase 10.1080/0001873031000093582} {\bibfield
  {journal} {\bibinfo  {journal} {Adv. Phys.}\ }\textbf {\bibinfo {volume}
  {52}},\ \bibinfo {pages} {219} (\bibinfo {year} {2003})}\BibitemShut
  {NoStop}%
\bibitem [{\citenamefont {Lubchenko}\ and\ \citenamefont
  {Wolynes}(2007)}]{Lubchenko2007}%
  \BibitemOpen
  \bibfield  {author} {\bibinfo {author} {\bibfnamefont {V.}~\bibnamefont
  {Lubchenko}}\ and\ \bibinfo {author} {\bibfnamefont {P.~G.}\ \bibnamefont
  {Wolynes}},\ }\href@noop {} {\bibfield  {journal} {\bibinfo  {journal} {Annu.
  Rev. Phys. Chem.}\ }\textbf {\bibinfo {volume} {58}},\ \bibinfo {pages} {235}
  (\bibinfo {year} {2007})}\BibitemShut {NoStop}%
\bibitem [{\citenamefont {Chandler}\ and\ \citenamefont
  {Garrahan}(2010)}]{Chandler2010}%
  \BibitemOpen
  \bibfield  {author} {\bibinfo {author} {\bibfnamefont {D.}~\bibnamefont
  {Chandler}}\ and\ \bibinfo {author} {\bibfnamefont {J.~P.}\ \bibnamefont
  {Garrahan}},\ }\href@noop {} {\bibfield  {journal} {\bibinfo  {journal}
  {Annu. Rev. Phys. Chem.}\ }\textbf {\bibinfo {volume} {61}},\ \bibinfo
  {pages} {191} (\bibinfo {year} {2010})}\BibitemShut {NoStop}%
\bibitem [{\citenamefont {Carleo}\ \emph {et~al.}(2012)\citenamefont {Carleo},
  \citenamefont {Becca}, \citenamefont {Schir{\'o}},\ and\ \citenamefont
  {Fabrizio}}]{Carleo2012}%
  \BibitemOpen
  \bibfield  {author} {\bibinfo {author} {\bibfnamefont {G.}~\bibnamefont
  {Carleo}}, \bibinfo {author} {\bibfnamefont {F.}~\bibnamefont {Becca}},
  \bibinfo {author} {\bibfnamefont {M.}~\bibnamefont {Schir{\'o}}}, \ and\
  \bibinfo {author} {\bibfnamefont {M.}~\bibnamefont {Fabrizio}},\ }\href
  {\doibase 10.1038/srep00243} {\bibfield  {journal} {\bibinfo  {journal}
  {Scientific reports}\ }\textbf {\bibinfo {volume} {2}},\ \bibinfo {pages}
  {243} (\bibinfo {year} {2012})}\BibitemShut {NoStop}%
\bibitem [{\citenamefont {van Horssen}\ \emph {et~al.}(2015)\citenamefont {van
  Horssen}, \citenamefont {Levi},\ and\ \citenamefont
  {Garrahan}}]{Horssen2015}%
  \BibitemOpen
  \bibfield  {author} {\bibinfo {author} {\bibfnamefont {M.}~\bibnamefont {van
  Horssen}}, \bibinfo {author} {\bibfnamefont {E.}~\bibnamefont {Levi}}, \ and\
  \bibinfo {author} {\bibfnamefont {J.~P.}\ \bibnamefont {Garrahan}},\ }\href
  {\doibase 10.1103/PhysRevB.92.100305} {\bibfield  {journal} {\bibinfo
  {journal} {Phys. Rev. B}\ }\textbf {\bibinfo {volume} {92}},\ \bibinfo
  {pages} {100305} (\bibinfo {year} {2015})}\BibitemShut {NoStop}%
\bibitem [{\citenamefont {Schiulaz}\ \emph {et~al.}(2015)\citenamefont
  {Schiulaz}, \citenamefont {Silva},\ and\ \citenamefont
  {M{\"u}ller}}]{Schiulaz2015}%
  \BibitemOpen
  \bibfield  {author} {\bibinfo {author} {\bibfnamefont {M.}~\bibnamefont
  {Schiulaz}}, \bibinfo {author} {\bibfnamefont {A.}~\bibnamefont {Silva}}, \
  and\ \bibinfo {author} {\bibfnamefont {M.}~\bibnamefont {M{\"u}ller}},\
  }\href {\doibase 10.1103/PhysRevB.91.184202} {\bibfield  {journal} {\bibinfo
  {journal} {Phys. Rev. B}\ }\textbf {\bibinfo {volume} {91}},\ \bibinfo
  {pages} {184202} (\bibinfo {year} {2015})}\BibitemShut {NoStop}%
\bibitem [{\citenamefont {Papi{\'c}}\ \emph {et~al.}(2015)\citenamefont
  {Papi{\'c}}, \citenamefont {Stoudenmire},\ and\ \citenamefont
  {Abanin}}]{Papic2015}%
  \BibitemOpen
  \bibfield  {author} {\bibinfo {author} {\bibfnamefont {Z.}~\bibnamefont
  {Papi{\'c}}}, \bibinfo {author} {\bibfnamefont {E.~M.}\ \bibnamefont
  {Stoudenmire}}, \ and\ \bibinfo {author} {\bibfnamefont {D.~A.}\ \bibnamefont
  {Abanin}},\ }\href@noop {} {\bibfield  {journal} {\bibinfo  {journal} {Ann.
  Phys.}\ }\textbf {\bibinfo {volume} {362}},\ \bibinfo {pages} {714} (\bibinfo
  {year} {2015})}\BibitemShut {NoStop}%
\bibitem [{\citenamefont {Barbiero}\ \emph {et~al.}(2015)\citenamefont
  {Barbiero}, \citenamefont {Menotti}, \citenamefont {Recati},\ and\
  \citenamefont {Santos}}]{Barbiero2015}%
  \BibitemOpen
  \bibfield  {author} {\bibinfo {author} {\bibfnamefont {L.}~\bibnamefont
  {Barbiero}}, \bibinfo {author} {\bibfnamefont {C.}~\bibnamefont {Menotti}},
  \bibinfo {author} {\bibfnamefont {A.}~\bibnamefont {Recati}}, \ and\ \bibinfo
  {author} {\bibfnamefont {L.}~\bibnamefont {Santos}},\ }\href {\doibase
  10.1103/PhysRevB.92.180406} {\bibfield  {journal} {\bibinfo  {journal} {Phys.
  Rev. B}\ }\textbf {\bibinfo {volume} {92}},\ \bibinfo {pages} {180406}
  (\bibinfo {year} {2015})}\BibitemShut {NoStop}%
\bibitem [{\citenamefont {Yao}\ \emph {et~al.}(2016)\citenamefont {Yao},
  \citenamefont {Laumann}, \citenamefont {Cirac}, \citenamefont {Lukin},\ and\
  \citenamefont {Moore}}]{Yao2016}%
  \BibitemOpen
  \bibfield  {author} {\bibinfo {author} {\bibfnamefont {N.~Y.}\ \bibnamefont
  {Yao}}, \bibinfo {author} {\bibfnamefont {C.~R.}\ \bibnamefont {Laumann}},
  \bibinfo {author} {\bibfnamefont {J.~I.}\ \bibnamefont {Cirac}}, \bibinfo
  {author} {\bibfnamefont {M.~D.}\ \bibnamefont {Lukin}}, \ and\ \bibinfo
  {author} {\bibfnamefont {J.~E.}\ \bibnamefont {Moore}},\ }\href {\doibase
  10.1103/PhysRevLett.117.240601} {\bibfield  {journal} {\bibinfo  {journal}
  {Phys. Rev. Lett.}\ }\textbf {\bibinfo {volume} {117}},\ \bibinfo {pages}
  {240601} (\bibinfo {year} {2016})}\BibitemShut {NoStop}%
\bibitem [{\citenamefont {Prem}\ \emph {et~al.}(2017)\citenamefont {Prem},
  \citenamefont {Haah},\ and\ \citenamefont {Nandkishore}}]{Prem2017}%
  \BibitemOpen
  \bibfield  {author} {\bibinfo {author} {\bibfnamefont {A.}~\bibnamefont
  {Prem}}, \bibinfo {author} {\bibfnamefont {J.}~\bibnamefont {Haah}}, \ and\
  \bibinfo {author} {\bibfnamefont {R.}~\bibnamefont {Nandkishore}},\ }\href
  {\doibase 10.1103/PhysRevB.95.155133} {\bibfield  {journal} {\bibinfo
  {journal} {Phys. Rev. B}\ }\textbf {\bibinfo {volume} {95}},\ \bibinfo
  {pages} {155133} (\bibinfo {year} {2017})}\BibitemShut {NoStop}%
\bibitem [{\citenamefont {Smith}\ \emph {et~al.}(2017)\citenamefont {Smith},
  \citenamefont {Knolle}, \citenamefont {Kovrizhin},\ and\ \citenamefont
  {Moessner}}]{Smith2017}%
  \BibitemOpen
  \bibfield  {author} {\bibinfo {author} {\bibfnamefont {A.}~\bibnamefont
  {Smith}}, \bibinfo {author} {\bibfnamefont {J.}~\bibnamefont {Knolle}},
  \bibinfo {author} {\bibfnamefont {D.~L.}\ \bibnamefont {Kovrizhin}}, \ and\
  \bibinfo {author} {\bibfnamefont {R.}~\bibnamefont {Moessner}},\ }\href
  {\doibase 10.1103/PhysRevLett.118.266601} {\bibfield  {journal} {\bibinfo
  {journal} {Phys. Rev. Lett.}\ }\textbf {\bibinfo {volume} {118}},\ \bibinfo
  {pages} {266601} (\bibinfo {year} {2017})}\BibitemShut {NoStop}%
\bibitem [{\citenamefont {Yarloo}\ \emph {et~al.}(2018)\citenamefont {Yarloo},
  \citenamefont {Langari},\ and\ \citenamefont {Vaezi}}]{Yarloo2018}%
  \BibitemOpen
  \bibfield  {author} {\bibinfo {author} {\bibfnamefont {H.}~\bibnamefont
  {Yarloo}}, \bibinfo {author} {\bibfnamefont {A.}~\bibnamefont {Langari}}, \
  and\ \bibinfo {author} {\bibfnamefont {A.}~\bibnamefont {Vaezi}},\ }\href
  {\doibase 10.1103/PhysRevB.97.054304} {\bibfield  {journal} {\bibinfo
  {journal} {Phys. Rev. B}\ }\textbf {\bibinfo {volume} {97}},\ \bibinfo
  {pages} {054304} (\bibinfo {year} {2018})}\BibitemShut {NoStop}%
\bibitem [{\citenamefont {Mondaini}\ and\ \citenamefont
  {Cai}(2017)}]{Mondaini2017}%
  \BibitemOpen
  \bibfield  {author} {\bibinfo {author} {\bibfnamefont {R.}~\bibnamefont
  {Mondaini}}\ and\ \bibinfo {author} {\bibfnamefont {Z.}~\bibnamefont {Cai}},\
  }\href {\doibase 10.1103/PhysRevB.96.035153} {\bibfield  {journal} {\bibinfo
  {journal} {Phys. Rev. B}\ }\textbf {\bibinfo {volume} {96}},\ \bibinfo
  {pages} {035153} (\bibinfo {year} {2017})}\BibitemShut {NoStop}%
\bibitem [{\citenamefont {Kob}\ and\ \citenamefont {Andersen}(1993)}]{Kob1993}%
  \BibitemOpen
  \bibfield  {author} {\bibinfo {author} {\bibfnamefont {W.}~\bibnamefont
  {Kob}}\ and\ \bibinfo {author} {\bibfnamefont {H.~C.}\ \bibnamefont
  {Andersen}},\ }\href@noop {} {\bibfield  {journal} {\bibinfo  {journal}
  {Phys. Rev. E}\ }\textbf {\bibinfo {volume} {48}},\ \bibinfo {pages} {4364}
  (\bibinfo {year} {1993})}\BibitemShut {NoStop}%
\bibitem [{\citenamefont {Jackle}\ and\ \citenamefont
  {Kronig}(1994)}]{Jackle1994}%
  \BibitemOpen
  \bibfield  {author} {\bibinfo {author} {\bibfnamefont {J.}~\bibnamefont
  {Jackle}}\ and\ \bibinfo {author} {\bibfnamefont {A.}~\bibnamefont
  {Kronig}},\ }\href@noop {} {\bibfield  {journal} {\bibinfo  {journal} {J.
  Phys. Condens. Matter}\ }\textbf {\bibinfo {volume} {6}},\ \bibinfo {pages}
  {7633} (\bibinfo {year} {1994})}\BibitemShut {NoStop}%
\bibitem [{\citenamefont {Pan}\ \emph {et~al.}(2005)\citenamefont {Pan},
  \citenamefont {Garrahan},\ and\ \citenamefont {Chandler}}]{Pan2005}%
  \BibitemOpen
  \bibfield  {author} {\bibinfo {author} {\bibfnamefont {A.}~\bibnamefont
  {Pan}}, \bibinfo {author} {\bibfnamefont {J.}~\bibnamefont {Garrahan}}, \
  and\ \bibinfo {author} {\bibfnamefont {D.}~\bibnamefont {Chandler}},\
  }\href@noop {} {\bibfield  {journal} {\bibinfo  {journal} {Phys. Rev. E}\
  }\textbf {\bibinfo {volume} {72}},\ \bibinfo {pages} {041106} (\bibinfo
  {year} {2005})}\BibitemShut {NoStop}%
\bibitem [{\citenamefont {Rokhsar}\ and\ \citenamefont
  {Kivelson}(1988)}]{Rokhsar1988}%
  \BibitemOpen
  \bibfield  {author} {\bibinfo {author} {\bibfnamefont {D.~S.}\ \bibnamefont
  {Rokhsar}}\ and\ \bibinfo {author} {\bibfnamefont {S.~A.}\ \bibnamefont
  {Kivelson}},\ }\href {\doibase 10.1103/PhysRevLett.61.2376} {\bibfield
  {journal} {\bibinfo  {journal} {Phys. Rev. Lett.}\ }\textbf {\bibinfo
  {volume} {61}},\ \bibinfo {pages} {2376} (\bibinfo {year}
  {1988})}\BibitemShut {NoStop}%
\bibitem [{\citenamefont {Moessner}\ and\ \citenamefont
  {Raman}(2011)}]{Moessner2011}%
  \BibitemOpen
  \bibfield  {author} {\bibinfo {author} {\bibfnamefont {R.}~\bibnamefont
  {Moessner}}\ and\ \bibinfo {author} {\bibfnamefont {K.~S.}\ \bibnamefont
  {Raman}},\ }in\ \href@noop {} {\emph {\bibinfo {booktitle} {Introduction to
  Frustrated Magnetism}}}\ (\bibinfo  {publisher} {Springer},\ \bibinfo {year}
  {2011})\ pp.\ \bibinfo {pages} {437--479}\BibitemShut {NoStop}%
\bibitem [{\citenamefont {Chalker}()}]{Chalker2017}%
  \BibitemOpen
  \bibfield  {author} {\bibinfo {author} {\bibfnamefont {J.~T.}\ \bibnamefont
  {Chalker}},\ }\href@noop {} {\bibinfo  {journal} {in {\it Topological Aspects
  of Condensed Matter Physics}, Lecture Notes of the Les Houches Summer School
  2014 edited by C. Chamon, M. Goerbig, R. Moessner, and L. Cugliandolo (Oxford
  University Press, 2017)}\ }\BibitemShut {NoStop}%
\bibitem [{Note1()}]{Note1}%
  \BibitemOpen
\bibfield  {journal} {  }\bibinfo {note} {The structure of $H_{\protect \text
  {QLG}}$ is similar to those in \cite {Horssen2015} and \cite {Hickey2016}.
  Constraints partition Hilbert space into disconnected components: states with
  only isolated vacancies cannot be dynamically connected with $H_{\protect
  \text {QLG}}$; but most states have at least one pair of neighouring
  vacancies and belong to the ergodic partition (we consider the dynamics in
  this main subspace). The model here and those of \cite
  {Horssen2015,Hickey2016} are termed ``embedded'' Hamiltonians in \cite
  {Shiraishi2017}.}\BibitemShut {Stop}%
\bibitem [{\citenamefont {Castelnovo}\ \emph {et~al.}(2005)\citenamefont
  {Castelnovo}, \citenamefont {Chamon}, \citenamefont {Mudry},\ and\
  \citenamefont {Pujol}}]{Castelnovo2005}%
  \BibitemOpen
  \bibfield  {author} {\bibinfo {author} {\bibfnamefont {C.}~\bibnamefont
  {Castelnovo}}, \bibinfo {author} {\bibfnamefont {C.}~\bibnamefont {Chamon}},
  \bibinfo {author} {\bibfnamefont {C.}~\bibnamefont {Mudry}}, \ and\ \bibinfo
  {author} {\bibfnamefont {P.}~\bibnamefont {Pujol}},\ }\href@noop {}
  {\bibfield  {journal} {\bibinfo  {journal} {Ann. Phys.}\ }\textbf {\bibinfo
  {volume} {318}},\ \bibinfo {pages} {316 } (\bibinfo {year}
  {2005})}\BibitemShut {NoStop}%
\bibitem [{\citenamefont {Lecomte}\ \emph {et~al.}(2007)\citenamefont
  {Lecomte}, \citenamefont {Appert-Rolland},\ and\ \citenamefont {van
  Wijland}}]{Lecomte2007}%
  \BibitemOpen
  \bibfield  {author} {\bibinfo {author} {\bibfnamefont {V.}~\bibnamefont
  {Lecomte}}, \bibinfo {author} {\bibfnamefont {C.}~\bibnamefont
  {Appert-Rolland}}, \ and\ \bibinfo {author} {\bibfnamefont {F.}~\bibnamefont
  {van Wijland}},\ }\href@noop {} {\bibfield  {journal} {\bibinfo  {journal}
  {J. Stat. Phys.}\ }\textbf {\bibinfo {volume} {127}},\ \bibinfo {pages} {51}
  (\bibinfo {year} {2007})}\BibitemShut {NoStop}%
\bibitem [{\citenamefont {Garrahan}\ \emph
  {et~al.}(2009{\natexlab{a}})\citenamefont {Garrahan}, \citenamefont {Jack},
  \citenamefont {Lecomte}, \citenamefont {Pitard}, \citenamefont {van
  Duijvendijk},\ and\ \citenamefont {van Wijland}}]{Garrahan2009}%
  \BibitemOpen
  \bibfield  {author} {\bibinfo {author} {\bibfnamefont {J.~P.}\ \bibnamefont
  {Garrahan}}, \bibinfo {author} {\bibfnamefont {R.~L.}\ \bibnamefont {Jack}},
  \bibinfo {author} {\bibfnamefont {V.}~\bibnamefont {Lecomte}}, \bibinfo
  {author} {\bibfnamefont {E.}~\bibnamefont {Pitard}}, \bibinfo {author}
  {\bibfnamefont {K.}~\bibnamefont {van Duijvendijk}}, \ and\ \bibinfo {author}
  {\bibfnamefont {F.}~\bibnamefont {van Wijland}},\ }\href@noop {} {\bibfield
  {journal} {\bibinfo  {journal} {J. Phys. A}\ }\textbf {\bibinfo {volume}
  {42}},\ \bibinfo {pages} {075007} (\bibinfo {year}
  {2009}{\natexlab{a}})}\BibitemShut {NoStop}%
\bibitem [{\citenamefont {Touchette}(2009)}]{Touchette2009}%
  \BibitemOpen
  \bibfield  {author} {\bibinfo {author} {\bibfnamefont {H.}~\bibnamefont
  {Touchette}},\ }\href@noop {} {\bibfield  {journal} {\bibinfo  {journal}
  {Phys. Rep.}\ }\textbf {\bibinfo {volume} {478}},\ \bibinfo {pages} {1}
  (\bibinfo {year} {2009})}\BibitemShut {NoStop}%
\bibitem [{SM()}]{SM}%
  \BibitemOpen
  \href@noop {} {}\bibinfo {howpublished} {See Supplemental Material for
  animations corresponding to Figures 3(d) and 5(c).}\BibitemShut {Stop}%
\bibitem [{\citenamefont {Lan}\ and\ \citenamefont {Powell}(2017)}]{Lan2017}%
  \BibitemOpen
  \bibfield  {author} {\bibinfo {author} {\bibfnamefont {Z.}~\bibnamefont
  {Lan}}\ and\ \bibinfo {author} {\bibfnamefont {S.}~\bibnamefont {Powell}},\
  }\href {\doibase 10.1103/PhysRevB.96.115140} {\bibfield  {journal} {\bibinfo
  {journal} {Phys. Rev. B}\ }\textbf {\bibinfo {volume} {96}},\ \bibinfo
  {pages} {115140} (\bibinfo {year} {2017})}\BibitemShut {NoStop}%
\bibitem [{\citenamefont {Garrahan}\ and\ \citenamefont
  {Newman}(2000)}]{Garrahan2000}%
  \BibitemOpen
  \bibfield  {author} {\bibinfo {author} {\bibfnamefont {J.}~\bibnamefont
  {Garrahan}}\ and\ \bibinfo {author} {\bibfnamefont {M.}~\bibnamefont
  {Newman}},\ }\href@noop {} {\bibfield  {journal} {\bibinfo  {journal} {Phys.
  Rev. E}\ }\textbf {\bibinfo {volume} {62}},\ \bibinfo {pages} {7670}
  (\bibinfo {year} {2000})}\BibitemShut {NoStop}%
\bibitem [{\citenamefont {Berges}\ \emph {et~al.}(2004)\citenamefont {Berges},
  \citenamefont {Bors\'anyi},\ and\ \citenamefont {Wetterich}}]{Berges2004}%
  \BibitemOpen
  \bibfield  {author} {\bibinfo {author} {\bibfnamefont {J.}~\bibnamefont
  {Berges}}, \bibinfo {author} {\bibfnamefont {S.}~\bibnamefont {Bors\'anyi}},
  \ and\ \bibinfo {author} {\bibfnamefont {C.}~\bibnamefont {Wetterich}},\
  }\href@noop {} {\bibfield  {journal} {\bibinfo  {journal} {Phys. Rev. Lett.}\
  }\textbf {\bibinfo {volume} {93}},\ \bibinfo {pages} {142002} (\bibinfo
  {year} {2004})}\BibitemShut {NoStop}%
\bibitem [{\citenamefont {Chamon}(2005)}]{Chamon2005}%
  \BibitemOpen
  \bibfield  {author} {\bibinfo {author} {\bibfnamefont {C.}~\bibnamefont
  {Chamon}},\ }\href@noop {} {\bibfield  {journal} {\bibinfo  {journal} {Phys.
  Rev. Lett.}\ }\textbf {\bibinfo {volume} {94}},\ \bibinfo {pages} {040402}
  (\bibinfo {year} {2005})}\BibitemShut {NoStop}%
\bibitem [{\citenamefont {Garrahan}\ \emph
  {et~al.}(2009{\natexlab{b}})\citenamefont {Garrahan}, \citenamefont
  {Stannard}, \citenamefont {Blunt},\ and\ \citenamefont
  {Beton}}]{Garrahan2009b}%
  \BibitemOpen
  \bibfield  {author} {\bibinfo {author} {\bibfnamefont {J.~P.}\ \bibnamefont
  {Garrahan}}, \bibinfo {author} {\bibfnamefont {A.}~\bibnamefont {Stannard}},
  \bibinfo {author} {\bibfnamefont {M.~O.}\ \bibnamefont {Blunt}}, \ and\
  \bibinfo {author} {\bibfnamefont {P.~H.}\ \bibnamefont {Beton}},\ }\href
  {http://www.pnas.org/content/early/2009/08/28/0902443106.abstract} {\bibfield
   {journal} {\bibinfo  {journal} {Proc. Natl. Acad. Sci. USA}\ }\textbf
  {\bibinfo {volume} {106}},\ \bibinfo {pages} {15209} (\bibinfo {year}
  {2009}{\natexlab{b}})}\BibitemShut {NoStop}%
\bibitem [{\citenamefont {Castelnovo}\ \emph {et~al.}(2012)\citenamefont
  {Castelnovo}, \citenamefont {Moessner},\ and\ \citenamefont
  {Sondhi}}]{Castelnovo2012}%
  \BibitemOpen
  \bibfield  {author} {\bibinfo {author} {\bibfnamefont {C.}~\bibnamefont
  {Castelnovo}}, \bibinfo {author} {\bibfnamefont {R.}~\bibnamefont
  {Moessner}}, \ and\ \bibinfo {author} {\bibfnamefont {S.}~\bibnamefont
  {Sondhi}},\ }\href@noop {} {\bibfield  {journal} {\bibinfo  {journal} {Annu.
  Rev. Condens. Matter Phys.}\ }\textbf {\bibinfo {volume} {3}},\ \bibinfo
  {pages} {35} (\bibinfo {year} {2012})}\BibitemShut {NoStop}%
\bibitem [{\citenamefont {Nandkishore}\ and\ \citenamefont
  {Hermele}(2018)}]{Nandkishore2018}%
  \BibitemOpen
  \bibfield  {author} {\bibinfo {author} {\bibfnamefont {R.~M.}\ \bibnamefont
  {Nandkishore}}\ and\ \bibinfo {author} {\bibfnamefont {M.}~\bibnamefont
  {Hermele}},\ }\href@noop {} {\bibfield  {journal} {\bibinfo  {journal}
  {arXiv:1803.11196}\ } (\bibinfo {year} {2018})}\BibitemShut {NoStop}%
\bibitem [{\citenamefont {Hickey}\ \emph {et~al.}(2016)\citenamefont {Hickey},
  \citenamefont {Genway},\ and\ \citenamefont {Garrahan}}]{Hickey2016}%
  \BibitemOpen
  \bibfield  {author} {\bibinfo {author} {\bibfnamefont {J.~M.}\ \bibnamefont
  {Hickey}}, \bibinfo {author} {\bibfnamefont {S.}~\bibnamefont {Genway}}, \
  and\ \bibinfo {author} {\bibfnamefont {J.~P.}\ \bibnamefont {Garrahan}},\
  }\href@noop {} {\bibfield  {journal} {\bibinfo  {journal} {J. Stat. Mech.}\
  ,\ \bibinfo {pages} {054047}} (\bibinfo {year} {2016})}\BibitemShut {NoStop}%
\bibitem [{\citenamefont {Shiraishi}\ and\ \citenamefont
  {Mori}(2017)}]{Shiraishi2017}%
  \BibitemOpen
  \bibfield  {author} {\bibinfo {author} {\bibfnamefont {N.}~\bibnamefont
  {Shiraishi}}\ and\ \bibinfo {author} {\bibfnamefont {T.}~\bibnamefont
  {Mori}},\ }\href {\doibase 10.1103/PhysRevLett.119.030601} {\bibfield
  {journal} {\bibinfo  {journal} {Phys. Rev. Lett.}\ }\textbf {\bibinfo
  {volume} {119}},\ \bibinfo {pages} {030601} (\bibinfo {year}
  {2017})}\BibitemShut {NoStop}%
\end{thebibliography}%

\end{document}